\RequirePackage{fix-cm}
\documentclass[twocolumn]{svjour3}          
\smartqed  
\usepackage{graphicx}
\usepackage{algorithm}
\usepackage[noend]{algpseudocode}
\usepackage{color}
\usepackage{xcolor}
\usepackage{algpseudocode,caption}
\usepackage{tcolorbox}

\usepackage{amsfonts,amsmath,graphicx,subfigure}
\algnewcommand\algorithmicswitch{\textbf{switch}}
\algnewcommand\algorithmiccase{\textbf{case}}
\algnewcommand\algorithmicassert{\texttt{assert}}
\algnewcommand\Assert[1]{\State \algorithmicassert(#1)}%
\algdef{SE}[SWITCH]{Switch}{EndSwitch}[1]{\algorithmicswitch\ #1\ \algorithmicdo}   {\algorithmicend\ \algorithmicswitch}%
\algdef{SE}[CASE]{Case}{EndCase}[1]{\algorithmiccase\ #1}{\algorithmicend\     \algorithmiccase}%
\algtext*{EndSwitch}%
\algtext*{EndCase}%
\algnewcommand\algorithmicleftcomment{$\triangleright $}
\algnewcommand\LeftComment{\State\algorithmicleftcomment\ }

\makeatletter
\newcommand\fs@plainruled{\def\@fs@cfont{\rmfamily}\let\@fs@capt\floatc@plainruled
  \def\@fs@pre{\hrule height.8pt depth0pt \kern2pt}%
  \def\@fs@post{}%
  \def\@fs@mid{\kern4pt\hrule height.8pt depth0pt \kern5pt}%
  \let\@fs@iftopcapt\iffalse}
\makeatother

\floatstyle{plainruled}
\restylefloat{algorithm}

\newtheorem{invariant}{Invariant}
\newtheorem{observation}{Observation}

\newcommand{\Alex}[1]{}
\newcommand{\Jon}[1]{}
\newcommand{\Cindy}[1]{}

\newcommand{\makealgtitle}{ {\vspace{-0.2cm}  \hrule height.8pt depth0pt \kern2pt}}
\newcommand{\nodeloc}{\beta}
\newcommand{\lcloc}{\alpha}
\newcommand{\nodeusr}{\gamma}

\newcommand{\DFRns}{\mbox{DFR}}
\newcommand{\DFR}{\mbox{DFR\ }}

\newcommand{\WStreamns}{\mbox{W-Stream}}
\newcommand{\WStream}{\mbox{W-Stream\ }}

\newcommand{\XS}{\mbox{XS\ }}

\newcommand{\XSCCns}{\mbox{XS-CC}}
\newcommand{\XSCC}{\mbox{XS-CC\ }}

\newcommand{\XStreamns}{\mbox{X-Stream}}
\newcommand{\XStream}{\mbox{X-Stream\ }}

\newcommand{\ufns}{\mbox{union-find}}
\newcommand{\uf}{\mbox{union-find\ }}

\newcommand{\bundlens}{\mbox{bundle}}
\newcommand{\bundle}{\mbox{bundle\ }}

\MakeRobust{\Call}

\begin{document}

\title{Connected Components for Infinite Graph Streams: Theory and Practice}


\author{Jonathan Berry \and Cynthia Phillips \and Alexandra\ Porter}


\institute{
           J. Berry \at
           Sandia National Laboratories\\
	\email{jberry@sandia.gov} 
	\and
	C. Phillips \at
           Sandia National Laboratories\\
	   \email{caphill@sandia.gov
        \and
       A. Porter \at
              Stanford University \\
              \email{amporter@cs.stanford.edu}           
          } 
}

\date{\today}

\maketitle

\begin{abstract}
Motivated by the properties of unending real-world cybersecurity streams, we present a new graph streaming model: \XStreamns. In this model, we maintain a streaming graph and its connected components at single-edge
granularity: every edge insertion could be followed by a query related to
connectivity.

In cybersecurity graph applications, the input stream typically consists of
edge insertions; individual deletions are not explicit.
Analysts will maintain as much history as possible and will trigger
customized bulk deletions when necessary for reasons related to storage
and/or systems management.
Despite a significant variety of dynamic graph processing systems and
some canonical literature on theoretical sliding-window graph streaming,
\XStream is the first
model explicitly designed to accommodate this usage model. Users can
provide boolean predicates to define bulk deletions, and \XStream will
operate through arbitrary numbers of applications of these events,
never filling up.  Queries are not allowed during
these bulk deletions, but edge arrivals are expected to occur continuously and
must always be handled.

\XStream is implemented via a ring of finite-memory processors. We give detailed algorithms for the system to maintain connected components on the input stream, answer queries about connectivity, and to perform bulk deletion. The system
requires bandwidth for internal messages that is some constant factor
greater than the stream arrival rate. 
 We prove a relationship among
four quantities: the proportion of query
downtime allowed, the proportion of edges that survive an 
aging event, the proportion of duplicated edges,
and the bandwidth expansion factor.

In addition to presenting the theory behind \XStreamns, we present
computational results for a single-threaded prototype implementation.
Stream ingestion rates are bounded by computer architecture.
We determine this bound for \XStream inter-process message-passing rates in
Intel TBB applications on Intel Sky Lake processors: between one
and five million graph edges per second.  Our compute-bound, single-threaded
prototype runs our full protocols through multiple aging events
at between one half and one a million edges per second, and we give
ideas for speeding this up by orders of magnitude.


\end{abstract}

\section{Introduction} \label{sec:intro}

\Cindy{Some stubs to include in the intro. We'll need to decide the extent this is a theory paper and the extent it is a systems/algorithms paper.  These pieces may move around when we turn our attention to the intro. I'm just putting down some thoughts from our discussions and some notation we will need as we go through the technical parts of the paper.}\Jon{I think this is mitigated now.  Cindy?}

We assume that the graph-edge stream is {\em effectively infinite}. This means that as long as the algorithm is running, it must always be prepared for the arrival of another edge. At any time, the system has seen only a finite set of edges and need store only a finite graph representation.  However this graph can be arbitrarily large and may eventually exceed any particular finite storage. In
order to deal with this case, we define a streaming model to exploit
distributed systems with huge aggregate memory and to handle bulk deletions
customized by a user-provided deletion or ``aging'' predicate.  The most
straightforward predicate would be $\Call{Age}{\mbox{t}}$ (get rid
of edges older than a certain timestamp), but users may require bulk
deletions guided by different predicates.  For example, they may wish to
preserve some old edges of high value.

Previous theoretical work on infinite graph streams in a sliding-window model
does allow for automatic time-based expiration while maintaining connected
 components~\cite{crouch2013dynamic,mcgregor2014graph}.  We adapt and
 generalize these theoretical ideas by allowing user-defined deletion
 predicates and distributed computation.  Furthermore, in
 Section~\ref{sec:related-work}
 we briefly survey previous work on dynamic graph processing.
 For now, we simply note that while some of this literature achieves impressive
 edge ingestion rates, none of it explains how to continue ingestion
 indefinitely as the storage capacity fills up.  These dynamic methods
 accept a stream of insertions and deletions, and if the former dominates the
 latter, the system will eventually fill and fail.  In this paper, we spend
 most of our effort providing a theoretical basis for graph stream
 computations of arbitrary duration.  We ensure that each edge is
 stored only once in our distributed model during normal conditions, and
 that we recover to that steady state in a predictable way after an
 aging event.

 Many dynamic graph processing systems ingest edges concurrently
 in large blocks, making it potentially impossible to detect the emergence and
 disappearance of fine-grained detail such as
 $O(\log n)$-sized components that merge into a giant component,
 as predicted by~\cite{chung}.  We model ingestion at a single-edge
 granularity to ensure that phenomena such as this will be observable.

\paragraph{Contributions} We give a distributed algorithm for maintaining streaming connected components. Processors are connected in a
one-way ring, with only one processor connected to the outside. The algorithm is designed for cybersecurity monitoring and has the following set of features:
\begin{itemize}
\item The system works with an arbitrary number of processors, allowing the system to keep an arbitrary number of edges.
\item Each edge is stored on only one processor, requiring $O(1)$ space, so the asymptotic space complexity is optimal.
\item The system fails because of space only if all processors are holding edges to their maximum capacity.
\item Processing an edge or query requires almost-constant time per system ``tick,'' the time to run union-find (inverse Ackermann's function).
\item Connectivity-related queries, spanning-tree queries, etc, are answered at perfect granularity. Though there is some latency, the answer is perfect with respect to the graph in the system at the time the query was asked.  This is in contrast to systems that process edge changes in batches up to millions allowing no finer granularity to queries.
\item Because some cyber phenomena do not have explicit edge deletions, the system removes edges only when required.
\begin{itemize}
\item This edge deletion is done in bulk.  Though querying is disabled during data structure repair, the system continues to ingest incoming edges. There is no need to buffer or drop edges if deletion happens during a period of lower use such as at night.
\item The analyst can select any (constant-time) edge predicate to determine which edges survive a bulk deletion.  This allows analysts to keep edges they feel are of high value regardless of their age.
\item For age-based deletion, the system can trigger and select correct parameters for bulk deletion automatically.
\end{itemize}
\item If the analyst selects legal values (depending on properties of the hardware and input stream) for how many edges survive a bulk deletion and what fraction of the time the system must answer queries, the system will run indefinitely.
\end{itemize}

\Cindy{TODO: What have I missed.  Also, a similar list for the experiments.}

\Cindy{These definitions should clarify a lot of concepts.  For example, a query sent at time t (that is between the tth potentially graph-changing event and the (t+1)st) answers the query based on graph $G(t)$.  It's possible for a vertex to be in $V(t)$ and $V(t'')$, but not $V(t')$ for some $t < t' < t''$.  At any time t, there is a primitive BB in the system for exactly each vertex in $V(t)$.}\Jon{This is mitigated, I think.  Cindy?}
\Cindy{NEW: I will check that we make each point.}

\section{Preliminaries}
\subsection{Modeling the graph through time}
We mark time based upon the arrival of any input stream element.  Time $t$
starts at zero and increments whenever a stream element arrives.
The input stream at time $t$ is the ordered stream that has arrived between
time $0$ and time $t$.
A stream element is an input edge $(u,v)$, a query (e.g., $\Call{Connected}{u,v}$), or a command (e.g. $\Call{Age}{timestamp}$).
In Section~\ref{sec:model} we formalize the operation
of our new model, \emph{\XStreamns}, at each of these ticks.
\begin{definition} 
\label{def:active}
\begin{tcolorbox}
\begin{description} 
\item \
\item[{\bf Active Edge:}] At any time $t > 0$, we say an edge is {\em active} if it has entered the system and no subsequent aging command has removed it. 
\item \
\item[{\bf Active Graph}:]
At any time $t > 0$, the active graph is $G_t = (V_t, E_t)$. Edge set $E_t$
is the set of active edges at time $t$ and the vertex set $V_t$ is the
set of endpoints of $E_t$.
\item \
\item[{\bf Active Stream}:] At any time $t > 0$, an active stream $A_t$ is
a subset of the input stream consisting only of active edges. \Cindy{We probably need a different notation, since we use $A_0$ to denote an initial finite graph input to DFR, following their notation.}
\end{description}
\end{tcolorbox}
\end{definition}

Note that $G_{t+1}$ differs from $G_t$ iff the stream element arriving at
time $t+1$ is an edge not already in $E_t$ or or an aging command.  In the
latter case, $E_{t+1}$ is the set of edges that survive the aging.

\noindent


\Cindy{The older, so far unrevised intro follows this point.}\Jon{mitigated, I think.  Cindy?}

\subsection{Streaming models}
In classic streaming models, computing systems receive data in a sequence of pieces and do not have space to store the entire data set. As in online algorithms, systems must make irreversible decisions without knowledge of future data arrivals. Graph streams are a type of such data streams in which a sequence of edges arrives at the computing system, which may assemble some of the edges into a graph data structure. Applications include modeling the spread of diseases in health-care networks, analysis of social networks, and security and anomaly detection in computer-network data. We focus on cybersecurity applications, in which analysts can infer interesting information from graphs that model relationships between entities. As the scale of such graphs increases, analysts will need algorithms to at least partially automate stream analysis.

We present detailed algorithms and a complete implementation of the real-time graph-mining methodology introduced in~\cite{AMP:berry2013maintaining}. In this streaming model, the full graph is stored in a distributed system. The model is also capable of bulk edge deletion, while continuing to accept new edges. The algorithm continuously maintains connected-component information.  It can answer queries about components and connectivity, except during a period of data-structure repair immediately following a bulk delete.

In classic graph streaming models, such as in~\cite{AMP:munro1980selection,AMP:muthukrishnan2005data,AMP:raghavan1999computing}, the input is a finite sequence of edges, each consisting of a pair of vertices. The edge sequence is an arbitrary permutation of graph edges, which may include duplicates. The output consists of each vertex, along with a label, such that two vertices have the same label if and only if they belong to the same component. Algorithms for the classic streaming model have two parameters: $p$, the number of times the algorithm can see the input stream, or the number of passes on the stream, and $s$, the storage space available to the algorithm. 

\Alex{need note about properties of algo, incl. asymptotically optimal space}\Jon{Mitigated, I think.  Alex?}
\Cindy{I didn't really see where we said that each edge requires O(1) storage across the whole system and therefore the space is optimal. I'm going to try to add a easy-to-find contributions section early in the introduction.}
\Jon{mitigated}

\begin{figure}[htb]
\begin{center}
\includegraphics[width=3.5in]{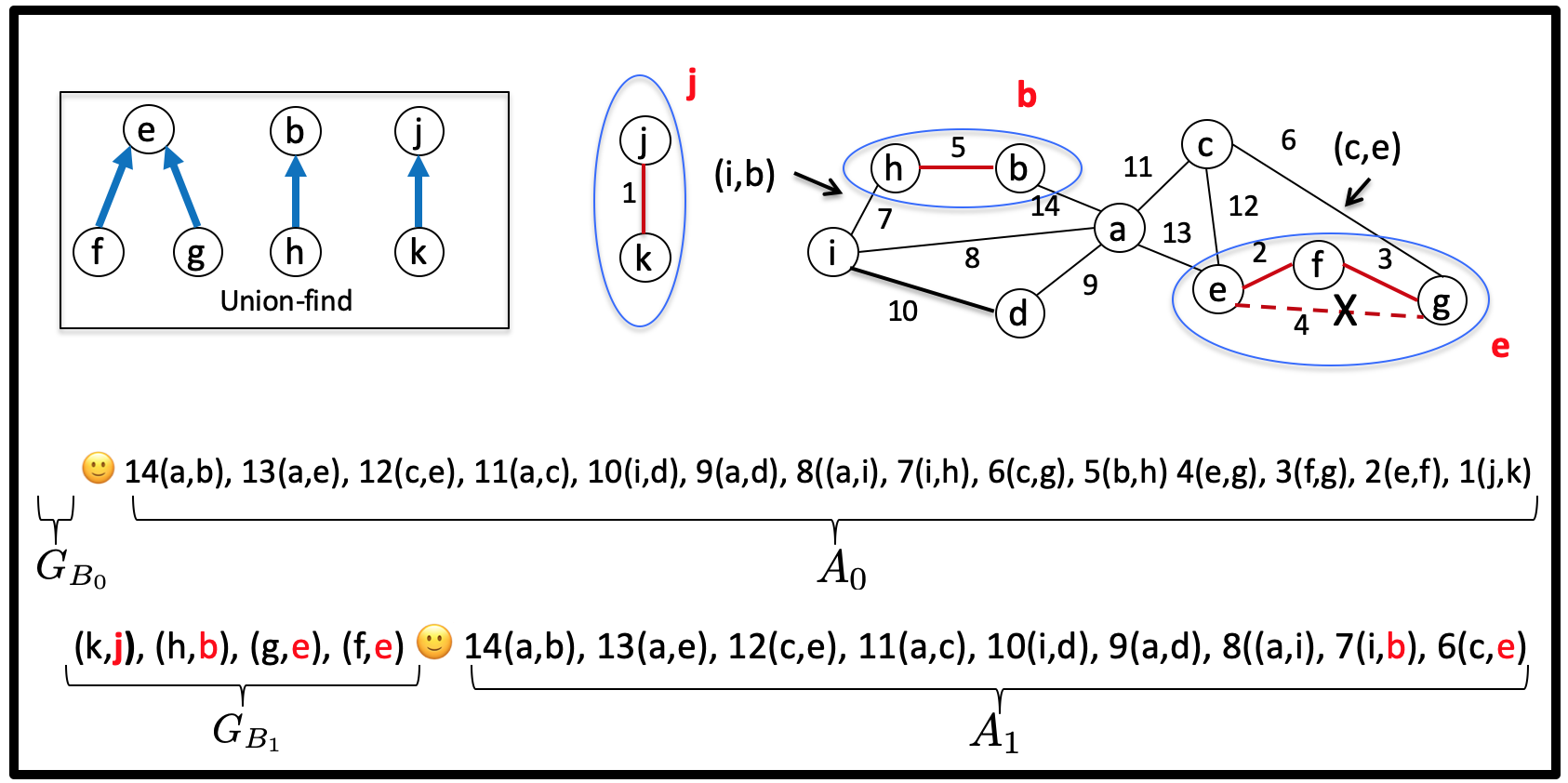} \\
\end{center}
\caption{\label{fig:wstream-example} Example of the first pass of the  \DFR 
connected components algorithm.
The \uf structure in the upper left has a capacity of four union
operations. This pass ingests five edges (shown in red) 
before filling, creating three super-nodes: $b$, $e$, and $j$,
which encapsulate vertices $f$, $g$, $h$, and $k$.
Edge $(e,g)$ was redundant. The remaining edges are relabled and
emitted as stream ``$A_1$,'' representing the contracted graph.
Then the contents of the \uf structure are emitted
(this is stream $B_1$).}
\end{figure}

The \emph{\WStreamns} model, developed in~\cite{AMP:demetrescu2009trading},
uses the concept multiple passes. Each pass can emit a customized
intermediate stream. \WStream can support several graph algorithms. However,
our work is specifically based upon the connected components algorithm
of~\cite{AMP:demetrescu2009trading}, which we will call \emph{\DFR} to
recognize the authors: Demetrescu, Finocchi, and Ribichini.

In the \emph{streamsort} model, introduced in~\cite{AMP:aggarwal2004streaming}, algorithms can create and process intermediate temporary streams, and reorder the contents of intermediate streams at no cost.

\subsection{\WStreamns}\label{sec:intro:wstream}
Because our work is an extension of the \DFR finite-stream connected-components algorithm based on the \WStream model, we describe that algorithm and model in detail now.  \DFR uses graph {\em contraction}. In contraction, one or more connected subgraphs are each replaced by a {\em supernode}.  Edges within a contracted subgraph disappear and edges with one endpoint inside a supernode now connect to the supernode.  For example, in Figure~\ref{fig:wstream-example}, connected subgraph (triangle) $f$, $e$, $g$, is contracted into supernode $e$.

In each pass, the W-Stream connected components algorithm ingests a finite stream and outputs a modified stream to read for the next pass. 
Each stream represents a progressively more contracted graph, where connected subgraphs contract to a node, until a single node represents each connected component. The stream at each pass comes in two parts. The first (A) part is the current, partially contracted graph, and the second (B) part lists the graph nodes buried inside supernodes. The initial input stream has all the graph edges in part A and an empty part B.  The last final-output stream has an empty part A and part B gives a component label for each vertex.
Figure~\ref{fig:wstream-example} illustrates the input and output of the first pass of the W-Stream connected-components algorithm.

During pass $i$, the algorithm ingests streams $A_{i-1}$ and $B_{i-1}$, in that order. 
First, it computes connected components using \uf
data structures until it runs out of memory.  More formally, $s$ is the
\emph{capacity} of the \WStream processor, measured in union operations. As shown in Figure~\ref{fig:wstream-example}, the set representative is one of the set elements, for reasons given later.
This \uf stage ingests a prefix of stream $A_{i-1}$.
Because its memory is now full, the processor must now emit information about the remaining stream rather than ingesting it. The algorithm incorporates what it has learned about the graph's connected components into the input for the next pass. Specifically, in the contracted graph corresponding to stream $A_i$ (called $G_{A_{i}}$), each set in the \uf data structure is represented as a supernode. 

The \DFR algorithm now generates the next stream $A_i$ from the remainder of stream $A_{i-1}$.
For each remaining edge $(u,v)$ in stream $A_{i-1}$, if endpoint $u$ is
buried within supernode $x$, we {\em relabel} $u$ to $x$. For example, 
in Figure~\ref{fig:wstream-example}, node $g$ is buried in supernode $e$.
Therefore, edge $(c,g)$ is relabeled to $(c,e)$.  If both endpoints are
relabeled to the same supernode, \DFR drops it according to contraction rules.

We now describe how to process stream $B_{i-1}$ and to emit stream $B_i$. In the first pass, stream $B_0$ is empty.  Stream $B_1$ tells which nodes are ``buried'' in the newly-created supernodes. Specifically, stream $B_1$ is a set of pairs $(u,x)$, where node $u$ is buried inside supernode $x$. Node $u$ will never appear in any future stream $A_i$. 
To process a non-empty stream $B_{i-1}$, we use the same relabeling strategy we used while emitting stream $A_i$.  However, the interpretation is different.  If $(u,x)$ is in stream $B_{i-1}$, and supernode $x$ is now part of a \uf set with representative $y$, we emit $(u,y)$ to stream $B_{i}$.  This means that node $u$ is part of supernode $y$ in stream $A_i$.

This process repeats until pass $f$ where $A_f$ is the empty stream. Stream $B_f$ can be interpreted as connected-component labels.  Two nodes have the same supernode label if and only if they are in the same connected component. 



We now summarize the argument from~\cite{AMP:demetrescu2009trading} that the \DFR algorithm is correct.  The $B_i$ streams are pairs of vertices from the original graph that are in the same connected component by correctness of the union-find algorithm.  We can therefore interpret the pair $(u,x)$ as an edge in a star-graph representation for a (partial) connected component (in their lingo, 
a ``collection of stars'').

Correctness of \DFR follows from maintaining the following invariant at each pass.
\begin{invariant} (\cite{AMP:demetrescu2009trading} Invariant 2.2)\label{inv:wstreamreqs}
For each $i \in \{0,..,p\}$, $G_{B_i}$ is a collection of stars and $G_{A_i} \cup G_{B_i}$ has the same connected components as $G$.
\end{invariant}

\begin{observation} 
\label{obs:wstreamreqs}
\DFR computes the same set of connected components for any permutation of
the input stream, and any arbitrary duplication of stream elements.
\end{observation} 

\Alex{stub:  each edge only stored once so space is linear; add note to intro about asymptotically optimal space?} \Cindy{This probably has to move earlier in the introduction.}\Jon{I think this is mitigated (see above).  Cindy and Alex?}

\begin{figure}[htb]
\begin{center}
\includegraphics[height=1.5in]{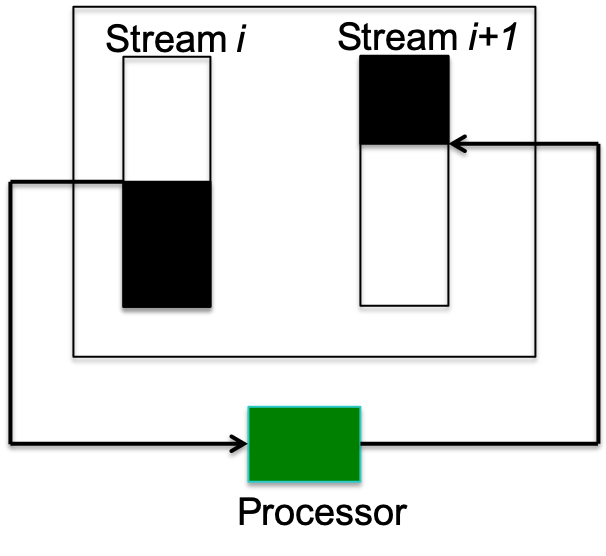} \\
\end{center}
\caption{\label{fig:wstream-impl} Implementation of the \DFR connected
components algorithm in the \WStream model.
The output of pass $i$ must be stored to disk, then read back in as input
for pass $i+1$.}
\end{figure}

\section{X-Stream model}\label{sec:model}
To motivate our \XStream model, we first consider how the theoretical
\WStream model might be implemented.  We show a plausible solution
in Figure~\ref{fig:wstream-impl}.  A single processor reads and writes
from files that store intermediate streams.  As these files may be of
arbitrary size, a direct implementation of \WStream is only a notional
idea.  \XStream is a theoretical variant of \WStream that can be
implemented efficiently.

In graph terms, \WStream stores only \emph{spanning tree} edges.  It
may drop any \emph{non-tree} edge since there is no concept of deletion
in that model.  Our \XStream model must accommodate bulk deletion by design
since
its input stream is infinite, and this means that non-tree edges must
be retained.  If a spanning-tree edge is deleted, then some non-tree edge
might reconnect the graph.

\begin{figure}[htb]
\centerline{\includegraphics[width=0.5\textwidth]{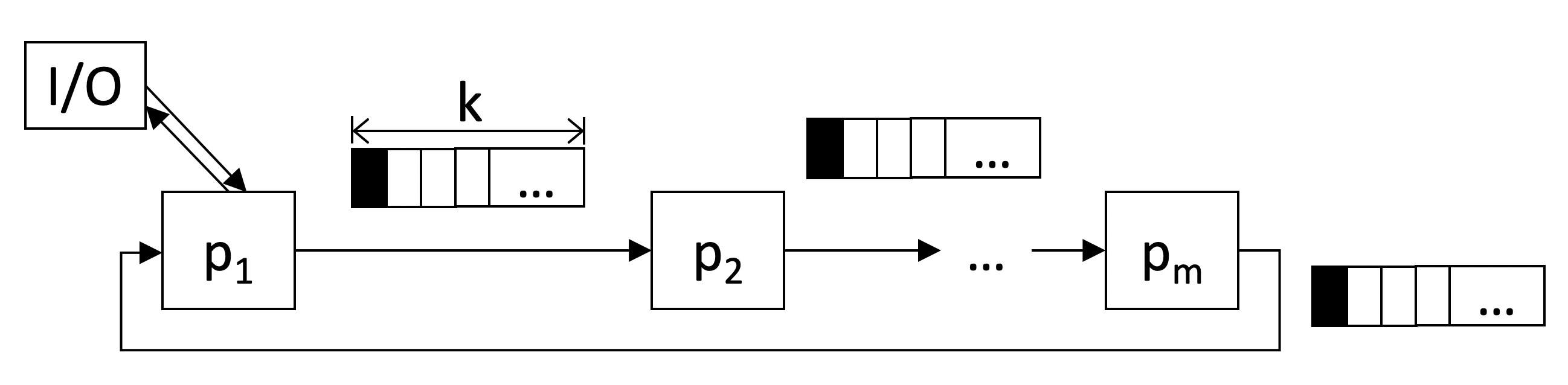}}
\caption{\label{AMP:fig:ringdiagram}\XStream architecture. Bundles of slots
circulate on a heartbeat. Input edges reside in the single primary (black) slot 
of each \bundlens. Payload (white) slots are used during bulk deletion and complex
queries.}
\end{figure}

Our \XStream model supports \WStreamns-like computations on infinite
streams of graph edges.  We present \XSCCns, a connected components algorithm
analogous to \DFR but implemented in the \XStream model.
Concurrent, distributed processing allows the \XSCC algorithm to handle streams without end markers, which are necessary in the \DFR algorithm to distinguish between stream passes. Aging also allows the \XStream model to manage unending edge streams in 
finite space.
\XStream has a ring of $p$ processors, plus an I/O processor, shown in Figure~\ref{AMP:fig:ringdiagram}. The I/O processor passes the input stream of edges, as well as queries and other external commands, to the first processor of the ring. The I/O processor also outputs information received from the ring, such as query responses.

Let ${\cal P} = \{p_0, \ldots, p_{P-1}\}$ be a set of non-I/O processors defining the
current state of the system. 
Processor $p_0$ is the \emph{head} (also called $p_H$), Processor
$p_{P-1}$ is the \emph{tail} (also called $p_T$), and we define 
successor and predecessor functions
as usual: $\mbox{pred}(p_i) = p_{j}$ for $j =i-1 \pmod{P}$, $\mbox{succ}(p_i) = p_{k}$ for $k = i+1\pmod{P}$. 
Thus each processor passes data to its successor, including the tail,
which passes data back to the head. 

Each $p_i$ has edge storage capacity $s_i$; for simplification we assume all processors have capacity $s$, for a total memory capacity of $S \equiv sP$ edges. We also assume $P \ll s$ since processors generally have at least megabytes of memory, even when there are enough processors for a relatively large $P$.

We next define a notion of time for the \XStream model. For this paper, we 
assume that all hashing and \ufns \ operations are effectively constant-time.

\vspace*{2mm}
\noindent
\begin{definition}[{\bf \XStream Step}] the \emph{\XStream clock} marks units of time, or \emph{ticks}. At each
tick, every processor is \emph{activated}: it receives a fixed-size \emph{\bundle}
of $k$ \emph{slots} from its predecessor, does a constant amount of
computation, and sends a size-$k$ output \bundle to its successor.
The head processor also receives a single slot of information from the I/O
processor at each tick.
\end{definition}

\XStream steps are thus conceptually systolic~\cite{kung1980algorithms}.
though real implementations such as the one we present in 
Section~\ref{sec:pseudocode} can be asynchronous.

\noindent
Using the notion of the \XStream clock, we can define the logical graph:
\begin{definition}
The {\em logical graph} at time $t$ is $G(t) = (V(t), E(t))$ is defined as:
\begin{enumerate}
\item the edge set $E(t)$ is the set of active edges at time $t$ and 
\item the vertex set $V(t)$ is the set of endpoints of $E(t)$.
\end{enumerate}
\end{definition}

Each message between ring processors is a \bundle of $k$ constant-sized slots. 
The constant $k$ is the \emph{bandwidth expansion factor}
mentioned in the abstract.  One of our key modeling assumptions is that
the bandwidth within a ring of processors is at least $k$ times greater than
the stream arrival rate, i.e., that the bandwidth expansion factor is at 
least $k$.  We give theory governing this factor in Section~\ref{sec:aging} 
and corroborate that theory via experiment in
Section~\ref{sec:experiments}.

Each slot can hold one edge and is
designated 
either \emph{primary} or \emph{payload} by its position in the \bundlens.  
By convention, the first slot in a \bundle is primary and all others are payload.
Primary slots generally contain information from outside the system, such as
input stream edges or queries, while payload slots are
used during the aging process and during queries with non-constant output
size (for example, enumerating all vertices in small components).

Once a processor receives a \bundle of slots, it processes the contents of each in turn. 
The processor can modify or replace the slot contents, as we will describe
below. When it has finished with all occupied slots, it emits the \bundle 
downstream. 


In order to formally define the graph stored in the \XStream model we consider
two \emph{edge states}:
A \emph{settled edge} is incorporated into the data structures of exactly one 
processor.
 A \emph{transit edge} is in the system, but is not settled. For example,
 it may still be moving from processor to processor in a \bundlens.


\subsection{Distributed Data Structures}
The \XStream data structure is a natural distributed version of the
W-Stream data structure, except that we must store the entire graph to allow
bulk deletions. \WStream pass $i$ identifies a set of connected subgraphs in
the input stream. \XStream processor $i$ stores a spanning forest of these
subgraphs.  By construction, the connected components of this forest are 
the same as those of the W-Stream connectivity structure, 
called $G_{B_i}$ in~\cite{AMP:demetrescu2009trading}.

\noindent
To describe how \XStreamns's distributed data structures implement the
W-Stream nesting strategy, we define the concepts of \emph{local
components} and \emph{building blocks}.
\begin{definition}
The connected components identified by the \uf structure on a processor $p_i$ are called the \emph{local components} (LCs) of $p_i$.
\end{definition}

\begin{figure}[hbt]
\centerline{\includegraphics[width=0.5\textwidth]{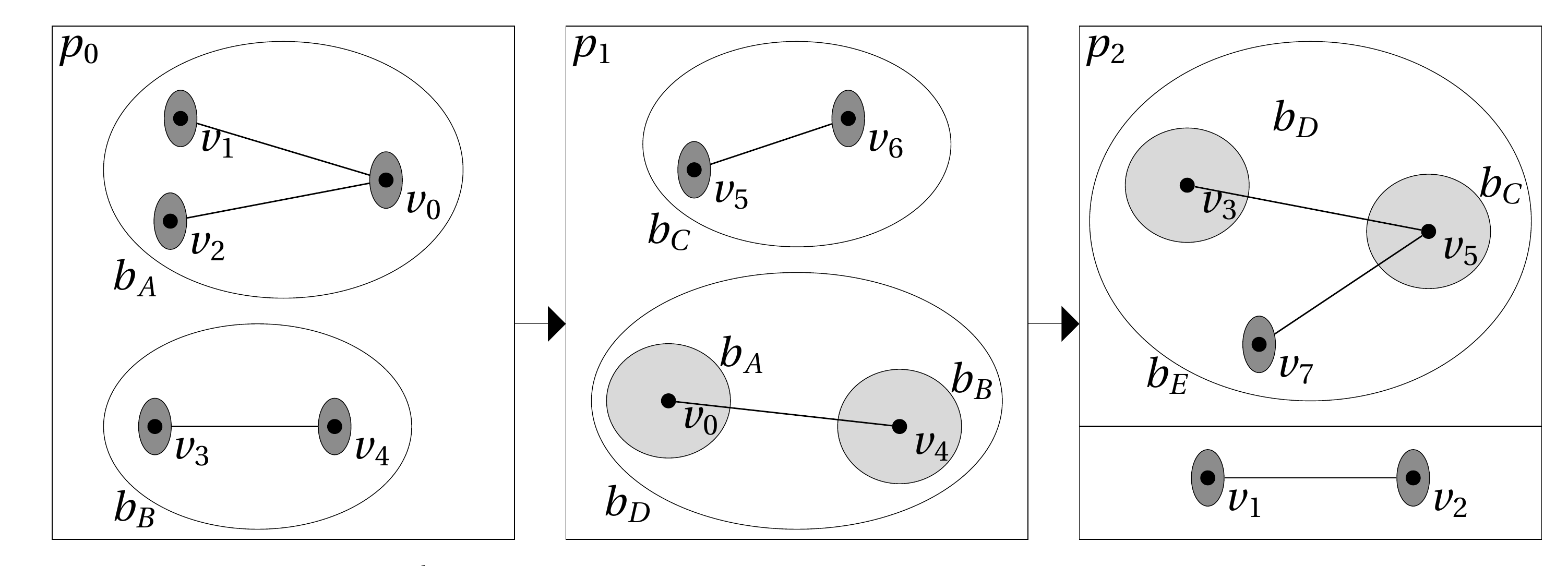}}
\caption{\label{fig:BBLCexample}
Example usage of notation: 
$\nodeloc(b_A)=p_1$ because processor $p_1$ consumed building block $b_A$, $\lcloc(b_A) = p_0$ because processor $p_0$ created local component $b_A$, $\nodeusr(b_A) = b_D$ because $b_A$ is a building block of local component $b_D$. Processor $p_1$ will relabel any vertex in $b_A$ to $b_D$ because
$\nodeloc(b_A)= p_1 = \lcloc(b_D)$. 
}
\end{figure}

A processor $p_j$ downstream of $p_i$ will see a local component of $p_i$ contracted into a single node, which $p_j$ might incorporate into one of its local components. Figure~\ref{fig:BBLCexample} shows an example of three processors and their \uf structures. As depicted in the first box of the figure, $b_A$ and $b_B$ are LCs of $p_0$ and $b_A$ and $b_B$ are incorporated in the LC $b_D$ on $p_1$.

\begin{definition}
 \emph{Building blocks} (BBs) for processor $p_j$ are the elements over which $p_j$ does \ufns.
 A \emph{primitive building block} contains exactly one vertex of the input graph. The set of all primitive building blocks is ${\cal B}_P$. A non-primitive building block corresponds to a local component from a processor upstream of $p_j$.
\end{definition}

We say a processor \emph{consumes} its building blocks because notification of
their existence arrives from upstream and does not propagate downstream.
A local component corresponding to a
set in a \uf structure \emph{encapsulates} all the building
blocks in that set.  We now formalize the relationship between \XStream
distributed data structures and \XStream processors.  Figure~\ref{fig:BBLCexample} illustrates these concepts.

\begin{definition} 
\label{def:alpha}
\begin{tcolorbox}
\begin{description} 
\item \
\item[{\bf $\lcloc$}:] The unique processor $p_i$ that creates local component $b_x$,  denoted $\lcloc(b_x) = p_i$.
\item[{\bf $\nodeusr$}:] the local component that contains building block $b_x$,
denoted $\nodeusr(b_x) = b_y$.
\item[{\bf $\nodeloc$:}]
The unique processor $p_j$ that consumes building block $b_x$, denoted
$\nodeloc(b_x) = p_j$.
\end{description}
\end{tcolorbox}
\end{definition}

\begin{definition}[\XStream nesting identity]
Let $b_x$ be a building block. Then
\begin{equation}
\lcloc(\nodeusr(b_x))  =  \nodeloc(b_x) \label{eq:identity}
\end{equation}
\end{definition}
\noindent
The \XStream Nesting Identity, which is true by construction, says that the
processor storing the local component that encapsulates building
block $b_x$ is the processor that consumed $b_x$.

We base our algorithm description and correctness arguments
on~(\ref{eq:identity}).  In the \XStream nesting structure, a building block $b_x$
is consumed by processor $\nodeusr(b_x)$ and incorporated into local
component $\nodeloc(b_x)$.  The latter is a creation event, and the
creating processor is $\lcloc(\nodeloc(b_x))$.

\begin{figure*}[thb]
\begin{center}
\includegraphics[width=5in]{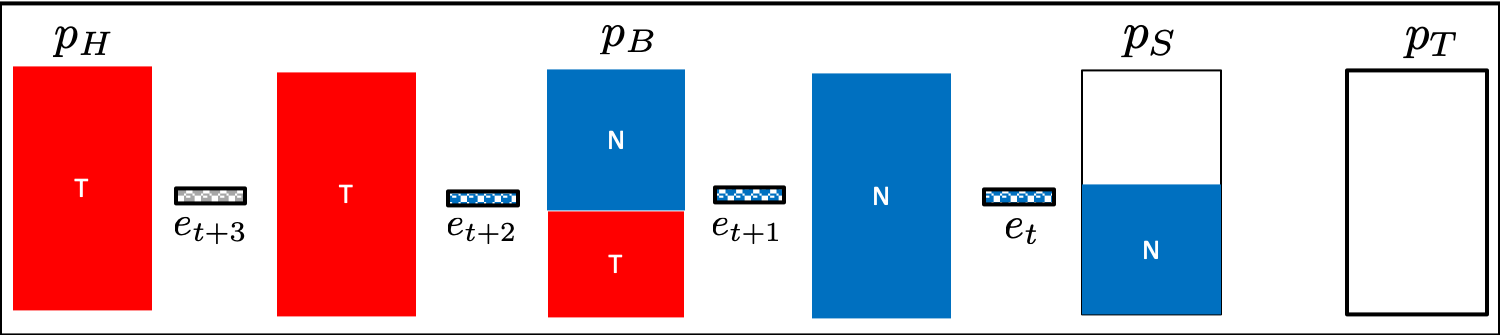}
\end{center}
\caption{\label{fig:xstream-normal} \XStream normal operation.  Spanning 
tree (red) edges are packed into $\{p_H,\ldots, p_B\}$ and store the 
connectivity information of W-Stream.  Non-tree (blue) edges are retained and
packed immediately afterwards. Note that $p_B$ will remain the building
processor until it has completely filled with tree edges (jettisoning its
current non-tree edges downstream). The gray edge ($e_{t+3}$) has not yet
been classified as tree or non-tree.  }
\end{figure*}

\subsection{Relabeling}
\label{sec:relabeling}
Suppose that $V$ is the domain of possible vertex names from the edge 
stream,  and ${\cal B}$ is the set of possible names of building blocks/local
components.
The \XStream \ nesting identity allows us to define a simple recursive \emph{relabeling function} $R_j : V \rightarrow {\cal B}$ for processor $p_j$, which returns the name of the local component that most recently encapsulates a primitive building block $b_v$ if such a local component exists, either on $p_j$ or upstream of $p_j$.  This function underlies correctness arguments
for \XStream data structures and connectivity queries.
 \begin{definition}
 Let the \emph{relabeling function} $R_j: V \rightarrow {\cal B}$ be defined as follows. For $v \in V$, let $R_{-1}(v) =b_z \in {\cal B}_P$, where $b_z$ is the primitive building block corresponding to vertex $v$.
For $j \geq 0$:
\begin{equation}
R_j(v) = \begin{cases}
         \nodeusr(R_{j-1}(v) )& \text{if $p_j=\nodeloc(R_{j-1}(v))$} \\
         R_{j-1}(v) & \text{otherwise}
	\end{cases}
\end{equation}
\end{definition}

In the example shown in Figure~\ref{fig:BBLCexample}, $R_1(v_1)= b_A$, $R_2(v_1) = b_D$, and $R_3(v_1) = b_E$. Since the first processor that knows of $v_5$ is $p_2$, $R_1(v_5) = R_{-1}(v_5)$, $R_2(v_5) = b_C$, and $R_3(v_5) = b_E$. $R_{-1}(v_1),R_{-1}(v_2),...,R_{-1}(v_7)$ are the primitive building blocks corresponding to the vertices and named using the vertex names. 

An edge $e$ is received by the system as two vertices with their primitive building blocks and a time stamp $t$: $e=([u,R_{-1}(u)],[v,R_{-1}(v)],t)$ and $R_{-1}(u),R_{-1}(v) \in {\cal B}_P$. Since edges are undirected, an edge \[e'=([v,R_{-1}(v)],[u,R_{-1}(u)],t')\] will be referred to as having the same endpoints as $e$. Processor $p_i$ for $i>0$ then receives edges in the form $e = ([u,R_{i-1}(u)],[v,R_{i-1}(v)],t)$. For all edges, the most recent timestamp is stored. 


\subsection{Operation Modes}

The \XSCC algorithm operates in two major modes. The \emph{aging} mode is active when a bulk deletion is occurring, as we will describe in Section~\ref{sec:aging}.  Otherwise the system is in \emph{normal} mode.

\XSCC uses the concept of relabeling, the \XStreamns \ ring
of processors, and periodic bulk deletion events to 
handle infinite streams of input edges. Each processor plays
the role of an intermediate stream in \WStreamns, borrowing the
concept of \emph{loop unrolling} from compiler optimization.  As
with the latter, we can store information concerning only a finite number of 
loop iterations (\WStream passes) at one time.  However, the periodic bulk 
deletions allow \XSCC to run indefinitely.

\section{Normal mode}\label{sec:normal}

During \XSCC normal mode, at any \XStream tick exactly one
processor is in a state of \emph{building}.
This building processor, $p_B$, accepts new edges, maintains connected
components with a \uf data structure, and stores spanning tree edges.
\XSCC normal mode maintains two key invariants, stated below and
illustrated in Figure~\ref{fig:xstream-normal}.
\begin{invariant}
Let $p_B$ be the building processor.  Then $p_i$ is completely full of
spanning tree edges for $0 \le i < B$ at all times, and $p_i$ has no
spanning tree edges for $i > B$. \label{inv:normal-1}
\end{invariant}

When $p_B$ fills with tree edges, a building ``token'' is passed downstream
to $p_B$'s successor, which assumes building responsibilities.  Thus, \XSCC
maintains a spanning forest of the input graph, packed into prefix processors
$\{p_0, \ldots, p_B\}$.
The \XSCC normal mode protocols maintain Invariant~\ref{inv:normal-1} and one other:

\begin{invariant}
Let $p_S$ be the first processor with any empty space.  Then $p_i$ is
completely full of edges for $0 \le i < S$ at all times, and $p_i$ has no
tree or non-tree edges for $i > S$. \label{inv:normal-2}
\end{invariant}


\setcounter{algorithm}{0}  
\begin{algorithm*}
\caption{This diagnostic routine is helpful for understanding correctness;
it would never be called in practice.  \label{algo:dump} 
This assumes the \WStream convention of choosing a vertex representative to name each 
supernode.
}
\XSCC diagnostic: dump connected components \\
\makealgtitle
\begin{algorithmic}[1]
\LeftComment{Precondition: the input stream has stopped. This never happens during normal operation.}
\LeftComment{Description: Processor $p_T$ emits correct finite stream connected components output.}
\Procedure{DumpComponentLabels}{$p_i$}
\LeftComment{Relabel all \Call{DumpComponentLabels}{} output from upstream}
          \While {\Call{Receive}{$u$, $R_{i-1}(u)$}}\label{lab:startRelabel}
\State        \Call{Emit}{$u$, $R_{i}(u)$}\label{lab:endRelabel}
          \EndWhile
\LeftComment{Now emit each union-find relationship}
          \For {$b_x : \nodeloc(b_x) = p_i$}   \Comment{$\nodeloc$ is the supernode relationship; see Definition~\ref{def:alpha}}\label{lab:startUFdump}
\State        \Call{Emit}{$b_x$, $R_{i}(b_x)$} \label{lab:endUFdump}
          \EndFor
\EndProcedure
\end{algorithmic}
\end{algorithm*}

Invariants ~\ref{inv:normal-1} and \ref{inv:normal-2} are illustrated in
Figure~\ref{fig:xstream-normal}, with sets of spanning tree edges represented
in red, and sets of non-tree edges represented in blue. In normal mode 
operation, single edges arrive at each X-Stream tick and propagate downstream
to the builder, being relabeled along the way. They settle into $p_B$ if 
they are
found to be tree edges, and into $p_S$ otherwise.  The figure
shows the system at \XStream tick $t+4$.  In this notional example, the
edge that arrived in the previous tick has passed through the head processor
$p_H$, but has not yet been resolved as ``tree'' or ``non-tree.''  Edge
$e_{t+2}$ has passed through two processors, and relabeling of the endpoints
has identified it as a non-tree edge. The basic protocol is thus quite simple;
the subtleties of \XSCC normal operation arise in maintaining the invariants. For example, the
builder may need to jettison non-tree edges downstream to make room for new tree edges.
We provide full detail in Section~\ref{sec:pseudocode}.

\subsection{\XSCC normal mode correctness}
We now show that Invariant~\ref{inv:normal-1} and
\XSCC relabeling implies an exact correspondence between
the connectivity structures computed by \XSCC and \DFRns.  

Algorithm~\ref{algo:dump} is a diagnostic routine intended to test
implementations of \XSCCns. At
\XStream time $t$, a call to this routine streams out the connected
components of the active graph $G_t$ as a stream of (vertex, label)
pairs.  Although we could correctly stream these $|V_t|$ vertex pairs
out even as new edges change the connected components (see
Section~\ref{sec:non-constant} for additional algorithm steps)\Jon{TODO: there
is nothing yet about this in Section~\ref{sec:non-constant}}, this version is more illustrative.
For simplicity we assume that the input stream pauses at time $t$.

When head processor $p_0$ receives the ``dump components'' command in a primary slot, it copies the command to the primary slot of the \bundle it will emit. 
Then, in Lines~\ref{lab:startUFdump}-\ref{lab:endUFdump} of Algorithm~\ref{algo:dump}, processor $p_0$ fills the remaining $k-1$ payload slots with relationships from its union-find structure, the way $\DFR$ outputs union-find information into the $B$ streams. Specifically, if $b_x$ is the name of a building block encapsulated by local component $b_y$ in $p_0$ (i.e. $\beta(b_x) = p_0$), then processor $p_0$ outputs a pair $(b_x, b_y = R_i(b_x))$, where $R_i(b_x)$ is the encapsulating supernode, the result of processor $p_0$ relabeling block $b_x$.  Processor $p_0$ then fills the payload slots of subsequent bundles until it has output all of its union-find relationships.

For downstream processor $p_i$ ($i > 0$), the \bundle that has the ``dump components'' command has $(u,R_{i-1}(u))$ pairs in the payload slots. In Lines~\ref{lab:startRelabel}-\ref{lab:endRelabel} of Algorithm~\ref{algo:dump}, processor $p_i$ relabels any block names $u$ that have been encapsulated by a new supernode in $p_i$. Otherwise $R_i(u) = R_{i-1}(u)$. After all relationships from upstream have arrived, there is empty payload for $p_i$ to output its union-find information as described above.

We now argue the correctness of \XSCC based on the correctness of
W-Stream. Let $D_i$ denote the output of processor $p_i$ from this diagnostic.  For the formal arguments, we require the following definitions:

\begin{definition}
During a union operation joining sets with representatives $b_x$ and $b_y$,
the \emph{supernode naming function} is $\eta: {\cal B} \times {\cal B} \rightarrow {\cal B}$ such
that $\eta(b_x,b_y)$ decides
whether $b_x$ or $b_y$ becomes the new set representative.
\end{definition}
For example, we
might choose a supernode naming function $\eta(b_x,b_y) = \min(b_x,b_y)$. This
is the function used in Figure~\ref{fig:wstream-example}.

\begin{definition}
\label{def:alg-with-params}
$\DFRns(s, \eta, A)$ is an implementation of \DFR with
processor union-find capacity $s$ and supernode naming function
$\eta$ run on input stream $A$. 
XS-CC$(s,\eta,A)$ is
defined similarly for XS-CC, with each processor's union-find
capacity set to $s$.
\end{definition}

A \emph{resolved} edge is one that has been classified as ``tree''
or ``non-tree.''\ Stream edges arrive
in an \emph{unresolved} state. 
In \DFRns, the stream $A_i$ written from pass $i$ contains only those edges that resolve to ``tree'' edges
(they connect supernodes in the current version of the contracted graph).  \DFR deletes as ``non-tree'' any edge that it determines to be contained inside a supernode.  
In contrast, \XSCC must retain all non-duplicate edges,
even after resolution.  In particular, non-tree edges
must be retained in case they are needed to reconnect pieces of the 
graph after bulk deletion. 
\XSCC removes duplicate edges from the stream after updating their 
timestamps.

By Invariant~\ref{inv:normal-1}, all unique non-tree edges (those contracted inside a supernode) are stored at the end of the \XSCC data structure in spare space in the builder, or in processors downstream of the builder. These downstream
processors have no
union-find structure. The following lemma ignores known non-tree edges.

\begin{lemma}
The stream of unresolved edges sent from processor $p_i$ to $p_{i+1}$ in 
$\XSCCns(s,\eta,A_0)$ is exactly $A_i$ from
$\DFRns(s,\eta,A_0)$.
\label{lem:A-stream}
\end{lemma}

\begin{proof}
We prove this lemma by induction.  For the base case, the first pass of \DFR and the first processor of \XSCC receive the same same finite stream of unresolved edges from the outside (logical processor $p_{-1}$), namely the input stream of edges $A_0$. Suppose that the stream of unresolved edges sent from processor $p_{i-1}$ to processor $p_i$ is the same as stream $A_{i-1}$ from \DFR.  We show that the stream of unresolved edges processor $p_i$ sends to $p_{i+1}$ is exactly \DFR stream $A_i$.

Processor $p_i$ of \XSCC and pass $i$ of \DFR begin by computing connected components via union-find.  Every edge that changes connectivity (starts a new component or joins two components) uses one of the $s$ possible union operations for this processor/pass. When they have both done $s$ union operations (their capacity), they have computed identical union-find data structures since they have done the same computations on the same input stream.  At this point, \DFR has not yet emitted any edges and \XSCC has emitted only resolved non-tree edges. Now \DFR processes the remaining edges of $A_{i-1}$, relabeling the endpoints, deleting edges where both endpoints are contained in the same supernode, and emitting the others to stream $A_i$. \XSCC relabels these remaining edges the same way, and emits the same stream of unresolved edges (among resolved non-tree edges).\qed
\end{proof}

Since \XSCC runs on unending streams, there is no ``end of stream'' mark to trigger creation of and processing of an \DFR $B_i$ stream.  However, the ``dump components'' diagnostic creates these $B$ streams.

\begin{lemma}
For $\XSCCns(s,\eta,A_0)$ followed by a call
to $\Call{DumpComponentLabels}{}$, stream $D_{i-1}$ is identical to
stream $B_i$ from $\DFRns(s,\eta,A_0)$.
\label{lemma:B-stream}
\end{lemma}
\begin{proof}
The ``dump components'' command after ingestion of a finite stream serves as an end-of-stream marker for \XSCC. We prove the lemma by induction. For the base case, streams $B_0$ and $D_{-1}$, the component information input to pass $0$ of \DFR and processor $p_0$ of \XSCC respectively are both empty. Suppose that stream $B_i$ from $\DFRns(s,\eta,A_0)$ is the same as stream $D_{i-1}$ from 
XS-CC$(s,\eta,A_0)$ followed
by a call to\\ $\Call{DumpComponentLabels}{}$. We show that stream $B_{i+1}$ is the same stream $D_i$. From the proof of Lemma~\ref{lem:A-stream}, the runs of \XSCC and \DFR compute the same connected components in processor $p_i$ and pass $i+1$ respectively. Because \XSCC and \DFR are using the same supernode naming function and have the same capacity, the union-find data structures (names of representatives and names of set elements) are identical. As described above, processor $p_{i}$ relabels and emits all elements of $D_{i-1}$ the same way that \DFR pass $i+1$ relabels elements stream $B_i$ to stream $B_{i+1}$. Then processor $p_i$ and \DFR pass $i+1$ output the information in their identical union-find structures in identical ways, completing streams $D_i$ and $B_{i+1}$ respectively. \qed
\end{proof}

\subsection{\XSCC queries}
The most basic \XSCC query is a connectivity query: are nodes $u$ and $v$ in
the same connected component? A query that arrives at \XStream tick $t$ will
be answered with respect to the graph $G_t$.  The query $(u,v)$ enters
the system from the I/O processor and propagates through the processors
just as new stream edges do.  Each processor relabels the endpoints,
and the tail processor returns ``$(u,v)$:yes'' if the labels are the same and ``$(u,v)$:no''
otherwise.  This holds even if  one or both of the endpoints have
never been seen before.
The following theorem shows that connectivity queries are correct at single-edge granularity, and therefore
that \XSCC in normal mode correctly computes the connected components of an
edge stream.

\begin{theorem}
Suppose that the connectivity query $(u,v)$ arrives at the head processor of an X-Stream
system with $P$ processors at X-Stream tick $t$. Then the I/O processor will
receive the boolean query answer at time $t+P$.  The answer will be \emph{True}
iff $u$ was connected to $v$ in $G_t$, the logical graph that existed at time $t$.
\label{thm:query-correctness}
\end{theorem}
\begin{proof}
Recall that $p_B$ is the building processor.
The query answer will be determined by
X-Stream tick $t+B$ at the latest, since, by Invariant~\ref{inv:normal-1},  
$p_B$ is the last
one to store any tree edges and hence, any union-find information.  Thus it is the 
last processor that can change a label. The $(u,v)$ query travels processor-to-processor in a primary basket, just as
the dump-components command does. If there are any transit edges in $G_t$ when the query arrives, they travel in slots of bundles strictly ahead of the query. Thus transit edges will settle into a processor before the query arrives.  Similarly, any edges that arrive after the query travel in bundles strictly behind the query and cannot affect the query relabeling. Thus when the bundle with the query arrives at processor $p_i$, the union-find data structure, and the processor's status as the builder or not, are set exactly according to the graph $G_t$ in the system when the query arrived.

Processing query $(u,v)$ is closely related to processing $\Call{DumpComponentLabels}{}$. Instead of dumping information for every vertex, starting at the point where a vertex is first encapsulated in a supernode, simple queries only consider two vertices. The label for $u$ will only change from $u$ to a supernode label $b_y$ at the processor $p_i$ that first incorporates $u$ into a local component ($p_i = \beta(u)$).  In $\Call{DumpComponentLabels}{}$, processor $p_i$ is the first that outputs any pair $(u, b_y)$, with first component $u$ into the stream $D_i$.  Thus, after the query has passed the building processor, the labels for vertices $u$ and $v$ are identical to their output values, which exit the system at time $t+P$. By Lemma~\ref{lemma:B-stream}, these are the same labels they would have if \DFR is run on graph $G_t$.  Because \DFR is a correct connected components algorithm, vertices $u$ and $v$ will have the same label if an only if they are in the same connected component. \qed
\end{proof}

We call queries that \XSCC answers with latency $p$ \emph{constant} queries.
See section~\ref{sec:non-constant} for examples of non-constant queries.

The next theorem shows that \XSCC is space-efficient, storing the current graph in asymptotically optimal space.
\begin{theorem}
\label{thm:non-dup}
In normal operation of \XSCCns, each edge is stored in exactly one processor,
requiring $O(1)$ space.
\end{theorem}
\begin{proof}
In normal operation, when a new edge $e$ arrives at a processor $p_i$ that
already stores a copy of $e$, processor $p_i$ removes $e$ from the stream and
updates the timestamp of $e$.
Invariant~\ref{inv:normal-1} ensures that incoming tree edge $e$ encounters any
previously-stored copies of itself before it reaches $p_B$, the building
processor, which recognizes it as a tree edge.
Invariant~\ref{inv:normal-2} ensures that incoming non-tree edge $e$
encounters any previously-stored copies of itself before it reaches $p_S$,
the first processor with any empty space. Furthermore, this invariant also
ensures that there are no edges stored downstream of $p_S$.  
\qed
\end{proof}

Theorem~\ref{thm:query-correctness} shows that basic connectivity queries are
answered correctly by \XSCCns.  In Section~\ref{sec:non-constant}
we informally discuss three additional types of feasible queries: complex queries such as
finding all vertices not in the giant component of a social network \Cindy{Can we ask for the maximum size of any connected component in a constant query? I think so, since every union-find component has a vertex count.  This gives us the size of the giant component.}\Jon{Last call for adding this.}, vertex-based
queries like finding the degree or neighborhood, and diagnostic
queries regarding system capacity used.  Also, by Invariant~\ref{inv:normal-1},
X-Stream always knows a spanning tree of the streaming graph by construction. This
tree could be checkpointed, for example, if processors share a filesystem. \Cindy{The information in the spanning tree is about the size of the output of dump components.}

\section{Aging mode}\label{sec:aging}
\begin{figure*}[bht]
\begin{center}
\begin{tabular}{c}
\includegraphics[width=5in]{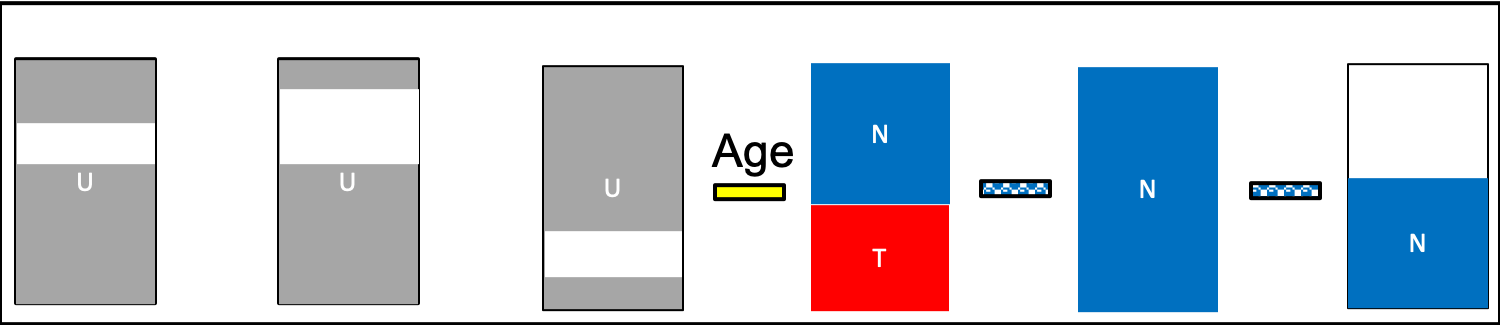} \\
(a) \\
\includegraphics[width=5in]{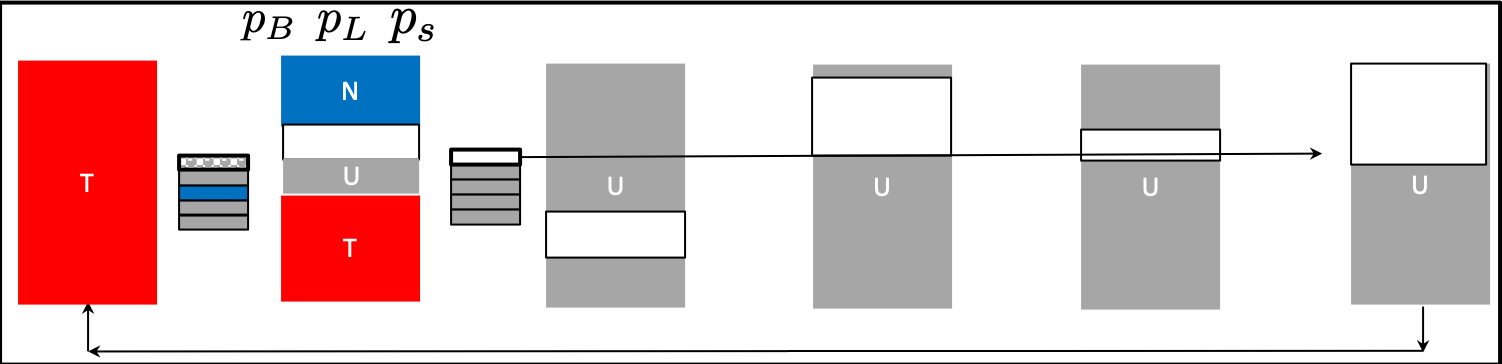} \\
(b) 
\end{tabular}
\end{center}
\caption{\label{fig:xstream-aging} \XSCC aging mode. (a) A token notifies processors
that they must apply an aging predicate.  All surviving edges become unresolved (gray).
(b) Normal operation continues uninterrupted for new edges, while
unresolved edges circulate back to be incorporated into a new data structure. Each processor
in turn becomes the loading processor ($p_L$) and recycles its unresolved edges.}
\end{figure*}

\begin{figure*}[thb]
\begin{center}
\includegraphics[width=5in]{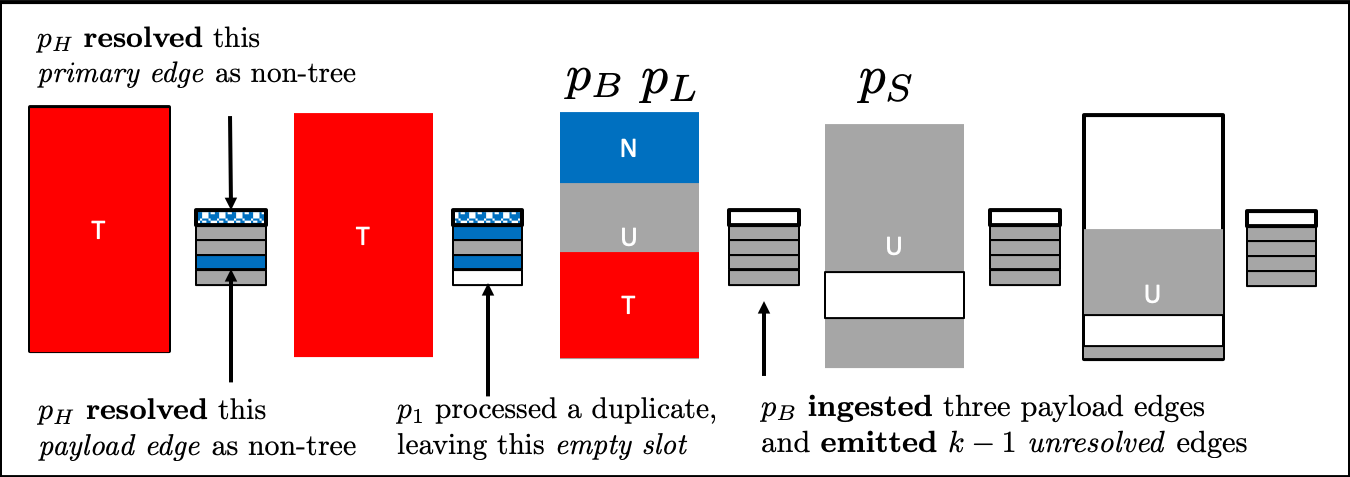} \\
\end{center}
\caption{\label{fig:xstream-aging-props} \XSCC aging nomenclature. both
primary
and payload edges are called \emph{resolved} when they have been
classified as tree or non-tree. Duplicate detection leaves empty
slots, and processors \emph{ingest} and \emph{emit} bundles of edges.}
\Jon{$p_H$ label, fix last ingest/emit example (my Office is currently broken;
will fix when Office fixed.}
\end{figure*}

\XSCC handles infinite streams via a bulk deletion operation we call
an \emph{aging} event.
Our model is thus unlike most previous work, in that we do not expect or
support individual
edge deletions embedded within the stream.  Rather, we expect the 
system administrator
to schedule bulk deletions to ensure that the oldest and/or 
least useful data are deleted in a timely manner.

To begin aging, the system administrator introduces an aging 
predicate (for example, a timestamp threshold) into the input stream.
The predicate
propagates through the system, and each processor suspends query 
processing upon receipt.  However, a new stream edge might arrive in the 
\XStream tick immediately after the aging predicate arrives from the 
I/O processor.
This and all other new edges must be ingested and processed without exception.  
Thus, the connectivity data structures must be rebuilt concurrently with
normal stream edge processing. When this rebuild is complete, queries
are accepted once again.

We now describe how \XSCC processes the aging predicate and prove correctness.
In Section~\ref{sec:aging-conditions} we provide theoretical guarantees relating 
the fraction of
system capacity used after the deletion predicate has been applied, the 
bandwidth expansion
factor, the proportion of query downtime that is tolerable, and the 
expected stream edge duplication rate.

\subsection{Aging process}
\label{sec:aging-process}
Figure~\ref{fig:xstream-aging} illustrates the aging process. An aging token arrives
with an edge-deletion predicate.  As the token propagates downstream, all
edges are reclassified to be \emph{untested}. If an edge later passes the
aging predicate it becomes \emph{unresolved} since the old connectivity
structure is no longer valid.
Immediately after the aging token is received
by the head processor, new stream edges may continue to arrive. These are
processed as normal, starting from empty data structures, so we maintain 
Invariants~\ref{inv:normal-1} and \ref{inv:normal-2} even during aging.

Conceptually, upon receipt of aging notification the deletion of all
edges that fail the aging predicate and
reclassification of all surviving edges to \emph{unresolved} is instantaneous.
However, in
practice each processor takes $\frac{s}{k-1}$ \XStream ticks to execute
a ``testing phase'' that applies the aging predicate to each stored
edge.  Without careful attention to detail,
implementers could allow a case in which there is no space yet for a new
stream edge.  In Section~\ref{sec:pseudocode} we give exact
specifications for a correct procedure that ensures no stream edge is
dropped, even in the \XStream tick immediately after aging notification.
If the testing phase has yet not identified empty space for a new stream
edge, then one of the unresolved edges can be sent downstream
in a primary slot.  This is an example of the jeopardy condition described
later in this section, corresponding to Line 21 in Algorithm~\ref{algo:pseudocode-driver}.

In addition to normal processing of new stream edges, \XSCC recycles all
unresolved edges that survive the aging predicate.
As depicted in Figure~\ref{fig:xstream-aging}, we introduce a new designation $p_L$
for a \emph{loading processor} or ``loader.''  Upon each activation to process
a stream edge, the loader packs unresolved edges into any available payload
slots in the output \bundlens.  Such bundles propagate around the ring. After
a bundle
reaches the head processor $p_H$, its payload edges are processed as if they
were new edges.  When the loader has emitted all of its unresolved edges, it
passes the loader token downstream to its successor.  Aging is complete when
the last processor with any unresolved edges has completed its loader duties.

\begin{figure*}[thb]
\begin{center}
\includegraphics[width=5in]{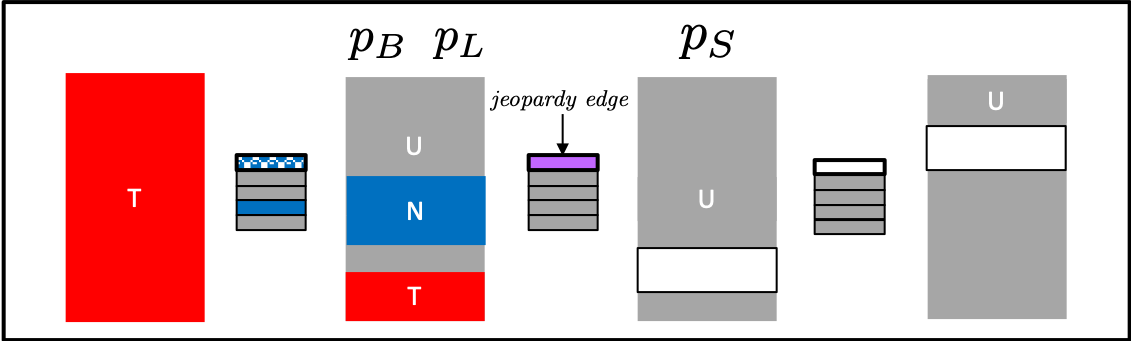} \\
\end{center}
\caption{\label{fig:jeopardy} The \XSCC aging ``jeopardy
condition.'' Processor $p_B$ currently bears both building and loading
responsibilities, is completely full of edges, and must ingest a \bundle
with no empty slots. It ingests $k$ slots, finds no duplicates,
and must emit $k$ slots. Therefore an unresolved ``jeopardy edge''
must be emitted in the primary slot.  If it doesn't settle in a 
processor before leaving the tail, the system is completely full and raises a
FAIL condition. Note that in this illustration, $p_S$ will be able to store
the jeopardy edge, so the jeopardy condition will soon be mitigated.}
\end{figure*}

The complete \XSCC protocols defined in Section~\ref{sec:pseudocode} 
enforce the previous invariants at all times, as well as the following
invariant during aging.  
\begin{invariant}
During aging, let $p_L$ be the loading processor and $p_B$ be the building processor. 
Then $B \le L$.  Also, processor $p_i$ has no unresolved edges for $i < L$
and $p_i$ has no resolved edges for $i > L$. \label{inv:aging-1}
\end{invariant}

The combination of all invariants ensures that all processors from the head to the builder are running \XSCC in normal mode on all incoming (and recycled) edges. All resolved edges are packed to the front (upstream).  When all edges have been recycled and aging ends, the layout of edges returns to normal mode.



Figure~\ref{fig:xstream-aging-props} puts the
nomenclature of our arguments into context. An edge becomes \emph{resolved}
when an XS-CC processor determines that it is a tree or non-tree edge,
regardless of whether it is a new stream edge in a primary slot or an
unresolved edge being recycled as payload.  Processors \emph{ingest} and
\emph{emit} bundles of edges.  With one exception we will discuss presently,
the complexity of processing input bundles and packing edges into
output bundles prior to emission is relegated to Section~\ref{sec:pseudocode}.

Aging is generally a straightforward process in which the loader token
steadily advances from $p_H$ to $p_T$, unresolved edges are recycled and
resolved, and the XS-CC connectivity structure is rebuilt.  
When builder and loader designations coincide in the same processor, 
that processor packs unresolved edges for emission first, then non-tree 
edges.  Edge bundles containing transit edges have one primary slot and $k-1$ 
payload slots, where
$k$ is the bandwidth expansion factor.  New stream edges reside in primary
slots, and unresolved edges circulate in payload slots until they are
resolved.  Payload edges continue in their assigned slots until allowed to 
settle, per the invariants.

There is a single exception to this last point, illustrated in
Figure~\ref{fig:jeopardy}. We call this the \emph{jeopardy condition}
and use it to specify exactly when the system fills
to capacity during aging (indicating that the aging command was too late
or did not remove enough edges).
In the jeopardy condition, processor $p_B$ is also the loader,
is already storing edges to its capacity $s$, and must ingest an edge \bundle
with no empty slots. It ingests $k$ slots, finds no duplicates,
and by conservation of space, must emit $k$ slots. 
Therefore an unresolved edge $e_j$ must reside in the primary slot.
If $e_j$ cannot be offloaded before exiting the tail,
the system is completely full and raises a FAIL condition.

The above discussion and the more detailed discussion in Section~\ref{sec:pseudocode} show the following property necessary for proving aging correctness holds:
\begin{property}
\label{prop:all-recycled}
During aging, every surviving edge is incorporated into the new 
connected components data structure either by $p_H$ directly or 
by traveling back to $p_H$ as a payload edge.
\end{property}

\subsection{Aging correctness}
We now argue correctness of the aging process.
We say that any implementation of \XSCC aging that maintains
Invariants~\ref{inv:normal-1}, \ref{inv:normal-2} and \ref{inv:aging-1} and property~\ref{prop:all-recycled} is \emph{compliant}.
A compliant aging process ensures that during aging there is a monotonic ordering of
edges in the system, with tree (red) edges never allowed downstream of
non-tree (blue) edges, and unresolved (gray) edges never allowed upstream
of non-tree edges. In the argument below, we slightly abuse notation by
using the graph $G_t$ in place of its edge set $E(G_t)$.

\begin{theorem}
Suppose a compliant XS-CC implementation receives an aging command at tick $t$ and reauthorizes queries at tick $t'$.  Let $F$ be the set of edges in $G_t$ that fail the aging predicate and let $E_{t\rightarrow t'}$ be the set of edges that arrive between time $t$ and $t'$. Then at tick $t'$, the x-stream system stores graph $G_{t'} = G_t - F \cup E_{t\rightarrow t'}$, can properly answer queries and stores each edge in $G_{t'}$ exactly once.
\label{thm:aging-correctness}
\end{theorem}
\begin{proof}
\Cindy{consistency in graph vs. edge set}\Jon{mitigated by text above. Cindy?}
As the aging command that arrived at time $t$ propagates through the
processors, they reclassify all current edges to ``untested'' as
described in Section~\ref{sec:aging-process}, forgetting the current
union-find structure. Thus the system starts processing a new graph
from an empty state at time $t+1$. As described in
Section~\ref{sec:aging-process}, processors delete all edges in $F$,
those who fail the predicate.  Each remaining edge in $G_t - F$ is
eventually loaded into payload slot by
Property~\ref{prop:all-recycled}, and processed at the head as
arriving edges. Invariants~\ref{inv:normal-1} and~\ref{inv:normal-2}
hold with the newly-created data structures thoughout aging.
Invariant~\ref{inv:aging-1} ensures that all unresolved edges are in
the builder processor or downstream. Those in the builder do not
affect the connectivity computation and are eventually moved
downstream. Thus, all edges arriving from outside the system are
processed as in normal mode and all edges arriving in the payload
slots are processed as in normal mode (other than traveling in a
payload slot).  Thus at time $t'$ when the tail processor passes the
loading token out of the system and enables queries, the \XStream
system stores exactly the edges in $G_t - F \cup E_{t\rightarrow t'}$,
with duplicates appropriately removed.  This is the graph the system
is required to hold by definition of aging and the requirement that it
drop no incoming edges during aging. The edges are processed into the
data structures with arbitrary mixing of new edges and recycled
(surviving) edges.  By Observation~\ref{obs:wstreamreqs}, and the equivalence of
\DFR and \XSCC in normal mode, the ordering of
the input edges does not matter for future query correctness.
By Theorem~\ref{thm:query-correctness}, the \XStream system will now correctly
answer querries on the graph starting at time $t'$.

During aging, some edges may be stored up to twice.  If a duplicate of a suriving edge $e$
enters the system before edge $e$ circulates back to the head processor, then edge $e$ is
stored both in the new data structure as a tree or non-tree edge and as an unresolved edge.
However, when edge $e$ is eventually recycled, it will be recognized as a duplicate and not
stored again.  By Theorem~\ref{thm:non-dup}, any edge that enters from outside the system during aging will be stored at most once in the new data structure.
\qed
\end{proof}

\section{Conditions for successful aging}
\label{sec:aging-conditions}
In this section, we define the conditions under which
a compliant aging process completes before the system fails for lack of space.
We consider properties of the system, properties of the input stream, and user
preferences.

\begin{tcolorbox}
\begin{definition}
\ \ \\
We define the following as tradeoff parameters associated with infinite runs of \XSCCns.
\label{def:infinite-run-params}
\begin{description}
\item[{\bf c:}] fraction of the total system storage occupied by
edges that survive the aging predicate
\item[{\bf d:}] percentage of \XStream ticks that the system is unavailable for queries due to aging
\item[{\bf u:}] estimate of the percentage of incoming stream edges that will be unique
\item[{\bf k:}] the bandwidth expansion factor: the size of an \XStream bundle (a set of edge-sized slots that circulates in the ring)
\item[{\bf p:}] number of \XStream processors
\item[{\bf S:}] aggregate storage available in the system
\item[{\bf s:}] storage per processor in a homogeneous system ($s = S/p$)
\end{description}
\end{definition}
\end{tcolorbox}

Aging must be initiated before the system becomes too full,
or else jeopardy edges will lead to a FAIL condition.  We quantify this
 decision point as follows.

\begin{lemma} \label{lemma:aging-lead-time}
In the worst case, there must be at least 
$$\frac{cS}{p(k-1)} + \frac{3}{2}p$$
open space in the system when an aging command is issued to be guaranteed sufficient space for aging, where
$c$, $S$, $p$ and $k$ are given in Definition~\ref{def:infinite-run-params}.  
\end{lemma}
\begin{proof}
When the aging command arrives, there could be $p$ edges in transit
that all must be stored. Because iteration over the untested list
doesn't imply
any specific ordering, in the worst case, when processors test the
edges against the predicate, all $cS$ surviving edges are tested
before an edge fails the predicate. This gives the latest time when
space becomes free for new edges. When a processor receives the aging
command, it processes $(k-1)$ untested edges each tick until it has
tested all its edges. In the $p$ ticks required for the aging command
to reach the tail, the head tests $p(k-1)$ edges, the second processor
tests $(p-1)(k-1)$ edges and so on, while the tail tests $(k-1)$
edges.  Thus in the first $p$ ticks after aging starts, the system tests
$\frac{p(1+p)(k-1)}{2}$ edges.  After that, the system tests $p(k-1)$
edges per tick. If the system tested $p(k-1)$, every tick, it would
require  $\frac{cS}{p(k-1)}$ ticks. But the first $p$ ticks are only half
as efficient, so we require an extra $p/2$ ticks. Thus the total
number of ticks before the system is guaranteed to remove an edge that
fails the predicate is at most $\frac{cS}{p(k-1)} + \frac{3}{2}p$.
\qed
\end{proof}

If the system is homogeneous, the empty space expression in Lemma~\ref{lemma:aging-lead-time} becomes $cs/(k-1) + \frac{3}{2}p$. For example, for a homogeneous system, assuming that $s \gg p$, if $c = 1/2$ and $k = 5$, then one should start aging while $1/8$ of the last processor is still empty. The last processor can issue a warning when it starts to fill and again closer to the deadline given $c$ and $k$.

\begin{theorem} \label{thm:infinite-runs}
In any \XSCC aging process initiated in accordance with Lemma~\ref{lemma:aging-lead-time}, if
$c$, $d$, $u$, $k$, $p$, $S$, and $s$ from Definition~\ref{def:infinite-run-params}
are set such that 
             \[k \ge   1 + \frac{(cp + 1) u}{dp (1-c)},\]
then the aging process will finish before the system storage fills completely.
\end{theorem}

\begin{proof}
After the aging token arrives, the head processor must apply the aging predicate
to its $s$ edges.  It processes $k-1$ per tick, as described in Section~\ref{sec:aging-process}.
Thus, after $\frac{s}{k-1}$ ticks, the head processor passes the loader token to the
second processor. By that time, all other processors have applied the predicate to
all of their edges and have a list of surviving edges.
Once unresolved edges begin circulating from
the loader (ignoring additive latencies such as the time until the first
payload reaches the head processor since $P \ll s$), $k-1$ edges re-enter the system to be
resolved at each
tick.  Since $cS$ unresolved edges survived the aging predicate, in the worst case
(when they are all in the second processor or later) it will take
$\frac{cS}{k-1}$ ticks to complete aging. During this time, every $\frac{1}{u}$
ticks yields a new, non-duplicate stream edge. Thus, the system will fill
to capacity in $\frac{1}{u} (1-c)S$ ticks.
The proportion $d$ constrains these two tick counts as follows:
\[ d \frac{1}{u} (1-c) S \ge \frac{cS + s}{k-1} = (cS / (k-1)) * (p+1)/p.\]
Simplifying this inequality and solving for $k$ (with Wolfram Alpha~\cite{wa},
for example)
yields the result. \qed
\end{proof}

The parameters $c$ and $d$ are user preferences, but $k$ is dictated by 
computer architecture.  Reasonable values of $k$ for current architectures are
$5-10$, but emerging data flow architectures may provide upward flexibility.
The parameter $u$ must be estimated by the user based on knowledge of the
input streams that s/he will feed to \XSCCns.

We can now state the central result of this paper. 

\begin{theorem}
\XSCC can process an effectively infinite stream of graph edges without failing,
answering
connectivity queries correctly when in normal mode, as long as
the system is configured in accordance with Theorem~\ref{thm:infinite-runs}
and aging is started with sufficient space available obeying
Lemma~\ref{lemma:aging-lead-time}.
\end{theorem}
\begin{proof}
Assuming that the proportion of \XStream ticks that yield a new,
non-duplicate stream edge is $u$,
an empty system will fill and fail in $\frac{1}{u}S$ ticks. Compliant
aging in accordance with Lemma~\ref{lemma:aging-lead-time} ensures that
aging will always complete before the system fills.  During normal
mode operation, Lemma~\ref{lemma:B-stream} and 
Theorem~\ref{thm:query-correctness} ensure, respectively, that 
accurate connected
component information is stored, and that connectivity queries are
answered correctly.  As long as the
system adminstrator adheres to such a schedule, \XSCC operation
can continue through an arbitrary number of aging events. \qed
\end{proof}
We note that queries yielding system capacity usage are TODO constant-size. In
the case of a simple aging predicate such as a timestamp threshold,
given a target proportion $c$ of edges that survive an aging
event, the \XStream system administrator could use an automated process to
trigger the aging process.

\section{X-Stream edge processing specification}\label{sec:pseudocode}
\setcounter{algorithm}{0}
\begin{algorithm*}
\caption{This is the driver function for an X-Stream implementation that is
compliant with Invariants~\ref{inv:normal-1}, \ref{inv:normal-2}, and
\ref{inv:aging-1}, and Property~\ref{prop:all-recycled}.
         \label{algo:pseudocode-driver}}
         X-Stream connected components driver \\
\makealgtitle
\begin{algorithmic}[1]

\Procedure{ProcessBundle}{\Call{PrimaryEdge}{e}, \Call{PayloadEdges}{$e_i$}}
\State  \Call{PackingSpaceAvailable}{} $= k$  \Comment{ the output buffer has size $k$ and is initially empty}
    \If {\Call{EmptyEdge}{$e$}}
\State  \Call{Pack}{EmptyEdge}  \Comment{any call to \Call{Pack}{} decrements \Call{PackingSpaceAvailable}{}}
    \Else \ \ \ \Call{ProcessEdge}{$e$}
    \EndIf
    \For {$e_i \in \Call{PayloadEdges}{}$} \Comment{there are payload edges only during aging or non-constant query processing}
\State  \Call{ProcessEdge}{$e_i$}
    \EndFor

\If {\Call{Not}{AGING}}
\State \Call{Emit}{PackedBundle}
\State \Return
\EndIf
\LeftComment{Aging-related logic}
    \If {\Call{Loader}{}}
        \If {\Call{Head}{}}  \Comment{the Head's testing \& resolution phase}
            \For {$i = 1, k-1$} 
\State          $e'$ = {\Call{PopEdge}{UNTESTED}} 
                \If {\Call{EmptyEdge}{$e'$}}
\State              \Call{Pack}{LoaderToken}
\State              \Call{Break}{}
                \EndIf
                \If {\Call{AgingPredicatePassed}{$e'$}}
                    \If {$i = k-1$ \mbox{and} \Call{Full}{}} 
\State                  \Call{Pack}{$e', \Call{Primary}{}$} \Comment{jeopardy edge}
                    \Else
\State                  \Call{ProcessEdge}{$e'$} \Comment{\Call{Head}{} immediately resolves surviving edge}
                    \EndIf
                \Else
\State              \Call{Delete}{$e'$}
                \EndIf
            \EndFor
        \Else   \Comment{ Downstream resolution phase (all testing has finished)}
            \For {$i \in 0, \ldots, \Call{PackingSpaceAvailable}{} - 1$}
\State          $e'$ = \Call{PopEdge}{UNRESOLVED}
                \If {\Call{NULL}{$e'$}}
\State              \Call{Pack}{LoaderToken}
\State              \Call{Break}{}
                \Else
\State              \Call{Pack}{e'}
                \EndIf
            \EndFor
        \EndIf
    \Else  \Comment{ Downstream testing phase}
        \If {\Call{Not}{HEAD} \Call{And}{} \Call{NotEmpty}{UNTESTED}}
            \For {$i = 1, k-1$} 
\State          $e'$ = {\Call{PopEdge}{UNTESTED}}
                \If {\Call{AgingPredicatePassed}{$e'$}}
\State              \Call{PushEdge}{$e'$, UNRESOLVED} 
                \Else
\State              \Call{Delete}{$e'$}
                \EndIf
            \EndFor
        \EndIf
    \EndIf
\State \Call{Emit}{PackedBundle}
\EndProcedure
\end{algorithmic}
\end{algorithm*}

Algorithms~\ref{algo:pseudocode-driver} and \ref{algo:pseudocode-functions}
show the \XSCC driver and constituent functions, respectively, for 
processing edges. We do not show full detail for token passes, commands,
and queries.
These functions maintain the invariants and produce a compliant XS-CC
implementation.  We used this pseudocode as guidance for the code
that produces our experimental results.

Each \XStream processor executes \Call{ProcessBundle}{} whenever it receives
the next bundle of edge slots, regardless of its current execution mode
(normal or aging).  It will process each slot in turn, and the 
constituent functions \Call{ProcessEdge}{}, \Call{ProcessPotentialTreeEdge}{},
and \Call{StoreOrForward}{} determine what to pack into an
output bundle destined to flow downstream.

Note that the top-level logic of processing the primary and payload
edges of a bundle is the same in Algorithm~\ref{algo:pseudocode-driver}, 
regardless of execution mode.  When a new
edge arrives from the stream, processors upstream of (and including) the 
building processor will classify it as tree or non-tree using the
relabeling logic of Section~\ref{sec:relabeling} (Lines 15-18 of 
\Call{ProcessEdge}{} and Lines 2-4 of \Call{ProcessPotentialTreeEdge}{}).
The builder stores any new tree edge. We ensure that this is
possible via logic to jettison an unresolved edge if one exists 
(only during aging; Lines 9 and 16 of \Call{StoreOrForward}{}), or else 
to jettison a non-tree edge (Line 15
of \Call{StoreOrForward}{}).  This progression of jettison logic maintains
Invariants~\ref{inv:normal-1} and \ref{inv:normal-2}.

Suppose that the head processor $p_H$ receives notification of an aging event
at \XStream tick $t$.  \XStream ticks $t$ and $t+1$ are especially interesting. 
If a new edge arrives in the input stream at $t+1$, it must be stored 
in $p_H$ (which
is now acting as both the builder $p_B$ and the loader $p_L$) in
order to maintain Invariant~\ref{inv:aging-1}.  However, $p_H$ has had only
one tick to initiate the process of testing its edges against the aging
predicate.  That means that it tested $k-1$ edges in tick $t$.  Suppose all
of these edges survived the predicate and therefore couldn't be deleted.
This is a jeopardy condition, and it was handled during tick $t$ by 
Lines 20-21 of 
\Call{ProcessBundle}{}.  Favoring the new edge, $p_H$ jettisoned in the
primary slot of its output bundle the last of the $k-1$ unresolved edges
it created in that tick.  Therefore, at tick $t+1$ we are assured that $p_H$ 
can store a new stream edge.

During aging, the loader $p_L$ packs unresolved edges into the empty payload
slots in incoming bundles to be sent around the ring.  When these edges
arrive at $p_H$, they are processed as if they were new stream edges,
classified as tree or non-tree, and incorporated into the data structures
in $\{p_H,\ldots,p_B\}$ by the same invariant-maintaining constituent 
functions that handle new edges.  One optimization we include is that
$p_H$ need not actually pack and send its unresolved edges around the
ring.  Rather, in Lines 13-23 of \Call{PackBundle}{}, $p_H$ simply tests
against the aging predicate and immediately processes its tested edges
rather than calling them unresolved.  

As aging proceeds, the \Call{Loader}{} token is passed downstream whenever
a processor exhausts its list of unresolved edges (Lines 28-31 of 
\Call{ProcessBundle}{}). Once the \Call{Loader}{} token exits the tail
processor, Property~\ref{prop:all-recycled} is established.

\begin{algorithm*}
\caption{These three constituent functions comprise the 
X-Stream algorithm for maintaining connected components.\label{algo:pseudocode-functions}}
         X-Stream constituent functions \\
\makealgtitle
\begin{algorithmic}[1]
\LeftComment{Processor $p_a$ receives an edge}
\Procedure{ProcessEdge}{$e=(u,v,R_{a-1}(u),R_{a-1}(v))$}
          \If {\Call{Duplicate}{e}} \Comment{regardless of $p_a$'s position in the chain, duplicate edges don't propagate downstream}
\State        \Call{SetNewestTimestamp}{e}  \Comment{During aging, either $e$ or its stored duplicate could be the newest}
              \If {\Call{Primary}{e}}
\State            \Call{Pack}{EmptyEdge} \Comment{bundles drive \XS ticks, \Call{ProcessBundle}{} requires a primary edge}
              \EndIf
\State        \Call{Return}{}
          \EndIf
          \If {\Call{DownstreamOfBuilder}{}} \Comment{$p_B \prec p_a$ : $p_a$ stores only non-tree and/or unresolved edges}
              \If {\Call{Primary}{e}} \Comment{need to store this edge if we can in order to ensure Invariant~\ref{inv:normal-2}}
\State            \Call{StoreOrForward}{e}        \Comment{\Call{StoreOrForward}{} accepts $e$ or packs it for output}
              \Else \Comment{\Call{Payload}{e}, i.e., aging}
\State            \Call{Pack}{e}         \Comment{processors downstream of the Builder simply propagate payload edges}
              \EndIf
\State        \Call{Return}{}
          \EndIf
\LeftComment{Processor $p_a$ contains connected component information, i.e., $p_a \prec = p_B$} 
          \If {$R_{a-1}(u) = R_{a-1}(v)$}   \Comment{previously-discovered non-tree edge}
\State        \Call{StoreOrForward}{e} 
          \ElsIf {\Call{Not}{\Call{ProcessPotentialTreeEdge}{e}}}\Comment{newly-discovered non-tree edge}
\State        \Call{StoreOrForward}{e}
          \EndIf
\EndProcedure
\end{algorithmic}

\begin{algorithmic}[1]
\Procedure{ProcessPotentialTreeEdge}{e=(u,v,\ldots)}
\State   $(R_a(u),R_a(v)) = $ \Call{Relabel}{e}
         \If {$R_a(u) = R_a(v)$}  \Comment{newly-discovered non-tree edge}
\State        \Call{Return}{FALSE}
         \EndIf
         \If {\Call{Builder}{}}      \Comment{builder $p_B$ must ingest tree edge $e$}
\State        \Call{Assert}{\Call{StoreOrForward}{e} = STORE} \Comment{ensure Invariant~\ref{inv:normal-1}}
              \If {\Call{FullOfTreeEdges}{}}
\State             \Call{Pack}{BuilderToken} \Comment{can be encoded with a bit; doesn't take a whole slot}
              \EndIf
         \Else \ \ \ \Call{Pack}{e}  \Comment{$p_a$ has previously sealed, so it is already full of tree edges}
         \EndIf
\State   \Call{Return}{TRUE}\Comment{still a potential tree edge; downstream processors will determine that}
\EndProcedure
\end{algorithmic}

\begin{algorithmic}[1]
\Procedure{StoreOrForward}{e=(u,v,\ldots)}
\LeftComment{Precondition: if $e$ is a tree edge, this processor is not full of TREE edges}
         \If {\Call{Full}{}}
             \If {\Call{Tail}{}}
\State            \Call{Fail}{} \Comment{the system is totally full}
             \EndIf
             \If {\Call{Unresolved}{$e$}}
\State                 \Call{Pack}{$e$}
\State                 \Call{Return}{FORWARD}
             \EndIf
\State       $e_p = $ \Call{PopEdge}{UNRESOLVED} \Comment{jettison an unresolved edge to keep a resolved one, if possible}
             \If {\Call{EmptyEdge}{$e_p$}} \Comment{no more edges to resolve}
                  \If {\Call{NonTree}{$e$}}
\State                 \Call{Pack}{$e$}  \Comment{no need to jettison a non-tree edge if $e$ is non-tree}
\State                 \Call{Return}{FORWARD}
                  \Else \Comment{by precondition, there must be a non-tree edge to jettison}
\State                 \Call{Pack}{\Call{PopEdge}{NONTREE}} \Comment{jettison a non-tree edge to keep a tree edge}
                  \EndIf
             \Else \ \ \ \Call{Pack}{$e_p$}
             \EndIf
         \EndIf
         \If{\Call{Primary}{e}} \Call{Pack}{\Call{EmptyEdge}{}}  \Comment{every output raft needs a primary edge}
         \EndIf
\State   \Call{Accept}{e}  \Comment{perform UNION/FIND if $\Call{Tree}{e}$}
\State   \Call{Return}{STORE}
\EndProcedure
\end{algorithmic}
\end{algorithm*}

\section{Related work}\label{sec:related-work}
This paper is motivated by the obseration that no other published work
meets the needs of the cybersecurity use case we describe in
Section~\ref{sec:intro}.  Most work on theoretical streaming problems,
best surveyed in~\cite{AMP:muthukrishnan2005data}, is limited to the finite case. Some research in the past decade has addressed 
infinite graph streaming in a sliding-window model~\cite{crouch2013dynamic,mcgregor2014graph} but the work is quite abstract, and expiration must be
addressed with every new stream observation.  We were unable to apply
this theory directly in a distributed system with bulk deletions, but our
\XSCC algorithm could be thought of as a generalized, distributed 
implementation of these sliding window ideas.

As far as we know, our \XStream model
and XS-CC graph-algorithmic use case comprise the first approach to
infinite graph streaming that is both theoretically-justified and practical.
We provided an initial view of the \XStream model in
a 2013 KDD workshop paper~\cite{AMP:berry2013maintaining}, and provide full
detail of a greatly-streamlined version in this paper. 

As the previous sections have made clear, our focus is infinite streaming of
graph edges with theoretical guarantees and a well-defined expiration 
strategy with a path to implementation in simple distributed systems.
Thus, we have approached the problem from a theoretical streaming perspective,
focusing primarily on related ``per-edge arrival'' streaming details.
We have shown how to maintain connectivity
and spanning tree information. We hope that others will expand the set
of graph queries available in \XStreamns, and/or propose new infinite
streaming models.

The closest related work comes from the discipline of \emph{dynamic graph
algorithms}, which takes a different approach.  Work in this area
typically assumes that the graph in its entirety is stored in a shared address
space or supercomputing resource.  Updates to the graph come in batches and 
take the form of edge
insertions and sometimes deletions too.  After a batch of updates is received,
incremental graph algorithms update attributes such as connected components
or centrality values.  During this algorithmic update the stream is typically
stopped.  There
is no attempt to describe what running infinitely in a finite space would mean
other than to rely on an implicit assumption that the batches will have as
many deletions as insertions over time.  We know that in the cybersecurity
context, for example, this assumption will never be true. An impressive
survey of dynamic graph processing systems is found in~\cite{besta2019practice}.

We break down the area of dynamic graphs into data structure work
that builds a graph without computing any algorithm
(for example~\cite{ediger2012stinger,riedy2011tracking,pearce1}),
work that ``stops the world'' at each algorithmic update
(for
example~\cite{riedy2011tracking,saga,helen}),
and recent attempts to
to process graph updates update algorithmic results concurrently~\cite{yin2018new,yin2019concurrent,sallinen2019incremental,grossman2020hoover}.

The data structure group includes solutions such as~\cite{ediger2012stinger},
 which achieves a rate of over 3 million edges per second on a Cray
 XMT2 supercomputer using a batch size of 1,000,000 edge updates,
 and~\cite{pearce1}, which achieves a rate of more than two billion edges
 per second on a more modern supercomputer while maintaining some
 information about
 vertex degrees.  While these rates are impressive, approaches such as these
 require a supercomputer and don't specify how to continue running as
 their storage fills up.

When incremental computation of graph algorithms such as Breadth-First
Search (BFS), connected components, PageRank, and others, SAGA-Bench~\cite{saga}
can achieve latencies of a fraction of a second on conventional hardware
using an update batch size of 500,000 edges.  This tranlates to a few
million updates per second, while also maintaining incremental graph
algorithm solutions.  Wheatman and Xu also exploit large batches of edge
updates and advanced data strctures (packed-memory arrays) to approach
this problem~\cite{helen}.  They achieve amortized update rates of up to
80 million updates per second while maintaining per-batch solutions to
graph problems such as connected components, where update batches can be of
size 10,000,000 or greater.  Even if our analysts could tolerate such
batch sizes, however, what prevents us from
simply adopting their approach is the requirement that we must
have a methodology for running infinitely.

We conclude our discussion of the dynamic graph literature with recent
results that process graph updates and update algorithmic results without
interrupting the input stream. The HOOVER system can run vertex-centric
graph computations on supercomputers that update connected components
information at an ingestion rate of more than 600,000,000 edges per
second~\cite{grossman2020hoover}.  However, the update algorithm works only
for edge insertions so our requirements are not met and the system would
 quickly fill up. Yin, et al.~\cite{yin2018new} propose a concurrent streaming
 model, and Yin \& Riedy~\cite{yin2019concurrent} instantiate this model with
 an experimental study on Katz centrality.  However, overlapping graph update
 and graph computation still does not meet our need for a strategy to compute
 on infinite streams.

\section{Experiments} \label{sec:experiments}

\begin{table*}[htb]
\begin{center}
\begin{tabular}{|c|c|c|c|c|} \hline
  & Bundle Size & 64-bit ints/s & \XStream potential ($k=2$) & \XStream potential ($k=5$) \\ \hline
Benchmark 1 & 5 & 1742160.27 & 174216.02 & 69686.41 \\ \hline
Benchmark 1 & 25 & 8680555.55 & 868055.55 & 347222.22 \\ \hline
Benchmark 1 & 250 & 54112554.11 & 5411255.41 & 2164502.16 \\ \hline
Benchmark 2 & 5  & 1344086.02 & 134408.60 & 53763.44 \\ \hline
Benchmark 2 & 25 & 6281407.03 & 628140.70 & 251256.28 \\ \hline
Benchmark 2 & 250 & 35063113.60 & 3506311.36 & {\bf 1402524.54} \\ \hline
\end{tabular}
\end{center}
\caption{\label{tab:benchmark} TBB benchmarks designed to produce bounds on \XStream performance
on Intel Sky Lake. Benchmark 1 propagates bundles downstream without any 
computation.  Benchmark 2 hashes two of every five integers in the bundle 
to simulate
\XSCCns's \Call{ProcessEdge}{} computation. The rightmost two columns show
upper bounds on \XSCC performance for bandwidth expansion factors $k=2$ and
$k=5$. On this architecture, we must send bundles of size 250 to maximize
performance. \XSCC with $k=5$ is bounded by 1.4 million edges per second.}
\end{table*}

The \XStream model and the \XSCC algorithm are based on message passing.
At each \XStream tick, each processor performs only a constant number of
operations.  These are predominantly hashing operations, union-find operations,
and simple array access.  Therefore, performance of \XSCC is strongly tied
to computer architecture. The faster a system can perform hashing and message
passing, the faster \XSCC will run.

With a current
Intel computer architecture (Sky Lake), we will show that our initial XS-CC
implementation can almost match the peak performance of a simple
Intel/Thread Building Blocks (TBB) benchmark that transfers
data between $P$ cores of the processor.
This translates to streaming rates of between half a million and one million 
edges per second,
which is comparable to the low end of performance spectrum for modern
dynamic graph solutions (none of which handle infinite streams).  The high
end of that spectrum is
not comparable to our context since we require no supercomputer and ingest data
from only one processor.  We have ideas to exploit properties of many
graphs (such as the phenomenon of a giant connected component)
for running many instances of \XSCC concurrently to boost our rates by
orders of magnitude.  However, that is beyond the scope of this paper.

\subsection{Computing setup and benchmarking}

All results in this paper were obtained using a computing cluster with
Intel Sky Lake Platinum 8160 processors, each with 2 sockets,
24 cores/socket, 2 HW threads/core (96 total), and 192GB DDR memory. The
memory bandwidth is 128GB/s, distributed over 6 DRAM channels.  The
interconnect is Intel OmniPath Gen-1 (100Gb/s).
The operating system is CentOS 7.9, and our codes are compiled with Intel icpc 
20.2.254 using the flags \verb+-O3 -xCORE-AVX512+.

Our full implementation of \XSCC is single-threaded~\footnote{the normal-mode
computations and data structures are single-threaded.  We use another thread
for cleanup and reallocation at an aging transition.} and
written in PHISH~\cite{phish}, a streaming
framework based on message passing.
However, before presenting XS-CC results,
we explore the expected peak performance of the algorihm on a single node of
the Sky Lake cluster using the vendor's own software library (Thread Building
Blocks 2019 8).

\paragraph{Mini benchmark}

We implemented a simple ring of X-Stream-style processing modules in TBB.
The head module
accepts bundles of synthetic data from an I/O module and sends them down the
ring toward the tail, which feeds back to the head. The latter merges this
bundle with its next input bundle.  We further distinguish two benchmarks:
\begin{itemize}
\item Benchmark 1: Each processor either simply copies input bundles to output.
\item Benchmark 2: Each processor hashes two of every five integer of the input bundle and copies the input to the output.  This approximately reflects the
main computation kernels of \XSCCns: hashing the timestamp of each each,
and doing a union-find operation.
\end{itemize}

Table~\ref{tab:benchmark} shows the performance of our TBB benchmarks as the
number of 64-bit integers in a bundle is varied. For these runs there are
10 processors in the ring.  Recall that \XSCC edges circulate as 5-tuples of
64-bit integers: $(v_1, l_1, v_2, l_2, t)$ where $v_i$ are vertex ids,
$l_i$ are local component labels, and $t$ is a timestamp.  Therefore the
raw rates of the third column must be divided by 5 to count in units of
\XStream primary edges.  Furthermore, to account for the payload slots in
\XSCC bundles active during aging or non-constant query processing,
the primary edge rate must be
divided by the bandwidth
expansion factor $k$. With optimal use of Intel's TBB, we see that we
should pass messages containing roughly 250 64-bit integers, and we
expect \XSCC edge streaming rates to be bounded by 1.4 million edges per
second.

Since Benchmark 2 is equivalent to Benchmark 1 except for a larger
compute load, we see clearly that Benchmark 2 is not bandwith bound 
on this 
architecture.  We believe that these benchmarks are bound by a combination
of compute and memory latency.  We experience a slow-down from 2.1 million 
edges per
second to 1.4 million simply by adding two hashing operations per bundle.
As our experiments with \XSCC will show, the latter is likely even more
compute bound.  This is welcome since it admits the possibility that
multithreading within TBB nodes and \XStream processors could accelerate our
single-threaded edge-processing results.

\begin{figure}[htb]
\begin{center}
\includegraphics[width=3in]{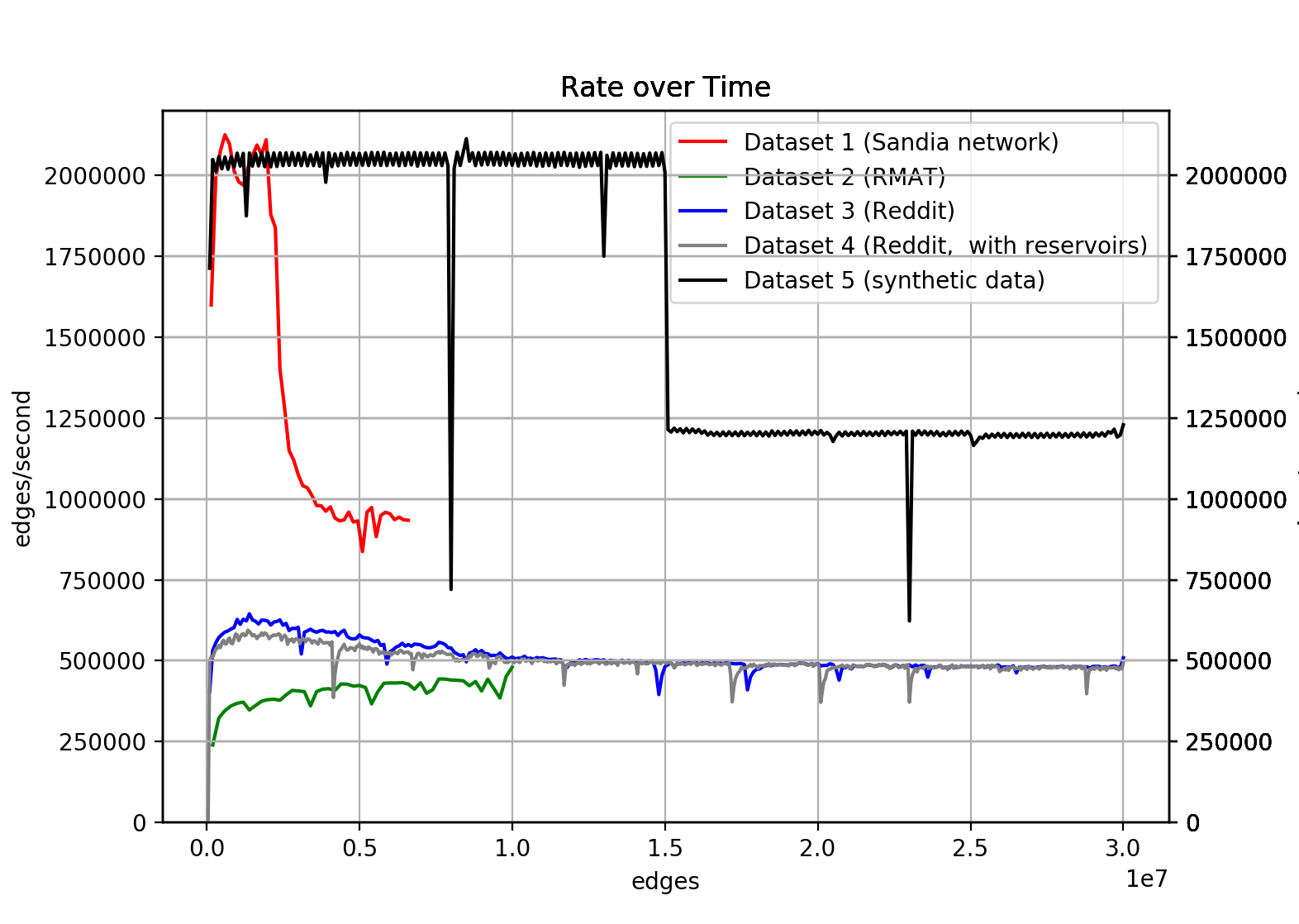}
\end{center}
\caption{Experiment 1: Prototype \XSCC normal-mode
streaming rates on the four datasets\label{fig:experiment1} on Sky Lake.
Table~\ref{tab:benchmark} places the expected peak performance of \XStream
computations on this architecture at between 1.4 million and 5 million edges
per second.  This indicates that our single-threaded \XSCC implementation is
not bandwidth bound and a future version could benefit from multithreading.
The decrease in performance from $\approx 2$M edges/sec to $\approx 1.25$M
edges/sec occurs when the second \XStream processor becomes the builder.}
\Alex{1=10m,2=reddit with reservoirs, 3=rmat, 4=synthetic (barbells) 5= reddit without reservoirs}
\end{figure}

\begin{figure}[htb]
\begin{center}
\includegraphics[width=3.5in]{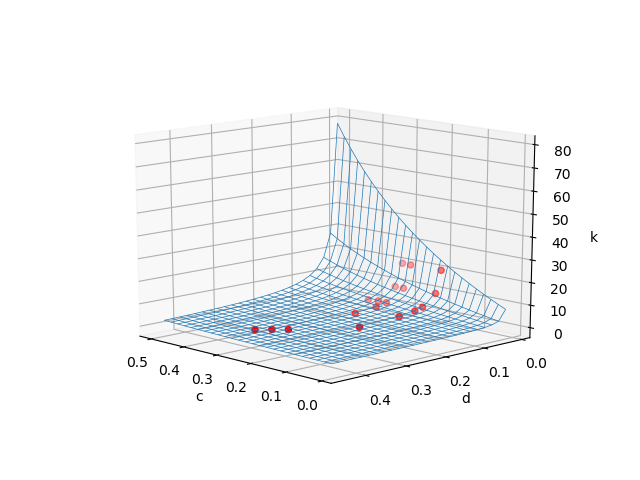}
\end{center}
\caption{Experiment 2: Empirical validation of Theorem~\ref{thm:infinite-runs}
using Dataset 3.  \label{fig:experiment2}  The value $c$ is the fraction of
edges that survive the aging predicate, the value $d$ is the fraction of 
time that queries are enabled, and the value $k$ is the bandwidth expansion
factor (the number of slots in a bundle).}
\end{figure}

Furthermore, in a real deployment of \XSCCns, we would assign a single
\XStream PE to a single compute node and communicate over the interconnect.
In fact, that is the basis of the \XSCC results presented in 
Figures~\ref{fig:experiment1}, \ref{fig:experiment2}, and \ref{fig:experiment3}.
In this case, we can compute the approximate theoretical peak for an
X-Stream-like computation as follows.  The interconnect is 100Gb/s, or
12.5GB/s. That translates to 1.5 billion 64-bit integers per second.  Since
\XSCC uses messages with 5 64-bit ints to represent an edge, and a typical
value of the \XStream bandwidth expansion factor $k$ is 5, we are bounded
by $1.5e9 / 5 / 5 \approx 62.5$ million \XSCC primary edges per second.
The rates of our prototype implementation do not approach this number, so
we believe that like the benchmarks, we are bound by a combination of
compute and memory latency.  A multi-threaded production version of
\XSCC would likely be necessary to better exploit a computing environment such
as our Sky Lake cluster.  With
that said: the TBB benchmark itself falls far short of the possible 
performance suggested by Sky Lake's theoretical peak memory bandwidth of 
128GB/s.
Significant algorithm engineering may be necessary to obtain a perfomant,
production version of \XSCCns.

\paragraph{Datasets}

We present prototype \XSCC results on three datasets:
\begin{enumerate}
    \item An anonymized stream of 10 million real gateway network traffic edges 
    from Sandia (the same stream used in ~\cite{AMP:berry2013maintaining}).
    \item A stream of edges from an R-MAT graph with 2097152 vertices,
    edge factor 8, and SSCA-2 parameters (0.45, 0.15, 0.15, 0.25)~\cite{rmat}.
    \item The Reddit reply network~\cite{kumar2018community} from SNAP~\cite{snapnets}, with 646,024,723 edges.
    \item A synthetic dataset with 100 continguous observations of
    each of a stream of edges with new, unique endpoints.
\end{enumerate}

For experiments below validating Theorem~\ref{thm:infinite-runs}, 
we note that Dataset 3 has a uniqueness parameter ($u$) of roughly 0.67.

\subsection{\XSCC implementation}
We used PHISH with the MPI back end to implement the \XSCC algorithm.  Stream
processing modules in PHISH are called ``minnows,'' and we instantiated a 
minnow to serve as the \XStream I/O processor and a group of minnows to form
the \XStream ring of procssors (one per compute node in the Sky Lake cluster).
We also ran with a single compute node hosting all \XSCC PE's.  However,
since our prototype is compute bound the rates we achieved were comparable
and are not presented.

We now present results from Sky Lake runs of our PHISH-based, single-threaded
implemenation of \XSCCns.  Before collecting these results, we
validated the correctness of \XSCC by designating every tenth stream
edge to be a connectivity query, statically computing the correct connected
components, and confirming that \XSCCns's query result matched that from
the static computation (with and without aging events).  We ran this
validation on a prefix of approximately 800,000 edges from
Dataset 3.

\subsection{Experiment 1: \XSCC normal-mode streaming rate}

Figure~\ref{fig:experiment1} shows \XSCC streaming rates for normal mode in
the three datasets.  We streamed the full Datasets 1 and 2 and a prefix of
30 million edges of Dataset 3.  Our single-threaded prototype implementation
is compute bound, as verified by computing to the benchmark results of
Table~\ref{tab:benchmark}.  Note that the performance of our prototype is
heavily data-dependent.  On the ``easy'' synthetic dataset (Dataset 5 in
the figure), note that we match rates with Table~\ref{tab:benchmark}.  When
real datasets cause more work, the ingestion rate drops, again showing that
we are compute bound.  Our prototype achieves rates between 500,000 and
1,000,000 edges per second, depending on the dataset. We note that Dataset 1,
which is a real dataset, has many repeat edges and admits an ingestion
rate of one million edges per second.

\begin{figure}[htb]
\begin{center}
\includegraphics[width=3.5in]{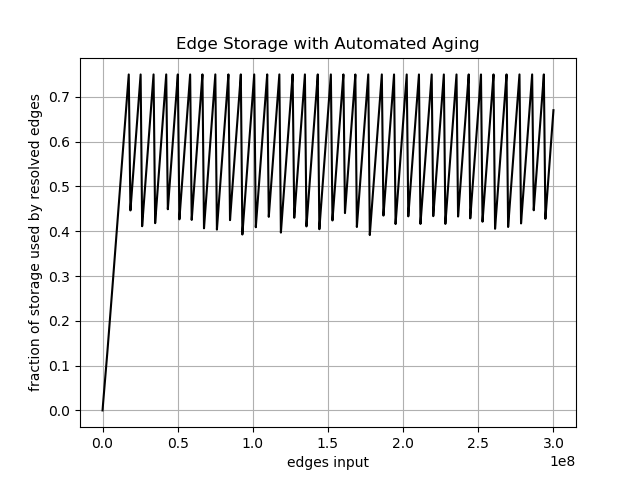}
\end{center}
\caption{Experiment 3: Automated aging with reservoir sampling on a
prefix of 300 million edges of Dataset 3.  The aging strategy
honors Theorem~\ref{lemma:aging-lead-time}.  This demonstrates
our strategy to run on infinite graph stream through arbitrary numbers
of aging events. \label{fig:experiment3}}

\end{figure}

\begin{figure}[htb]
\begin{center}
\includegraphics[width=3.5in]{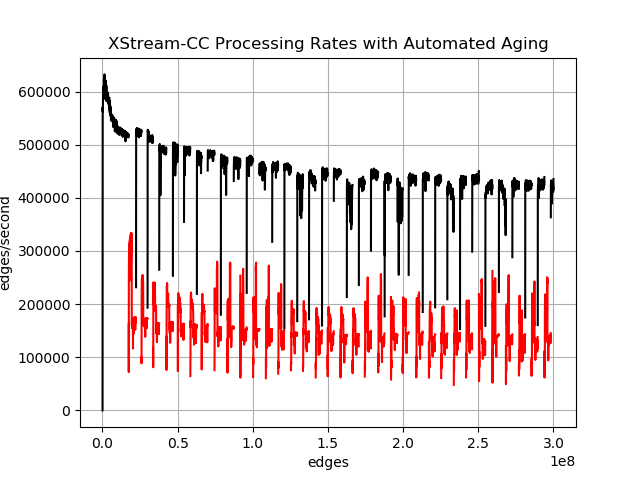}
\end{center}
\caption{Experiment 3: The stream processing rate of our
prototype over time during
automated aging.\label{fig:experiment3-rates}.  
Although ingestion rate of new edges does slow during aging due
to factors not modeled such as the proportion of tree edges,
we note that aging during periods of lesser stream intensity 
(such as nighttime for a worksite) should allow infinite 
streaming without dropping edges.
    \Alex{it appears to be because we have more tree edges over time, working on verifying that is correct (treetestrates\_10kto10m.png shows how rate slows down as tree edges are a bigger and bigger fraction, from 10k out of 10m in red to 10m out of 10m in gray)}
}
\end{figure}

\subsection{Experiment 2: \XSCC with a single aging event}
Recall that Theorem~\ref{thm:infinite-runs} relates the \XStream bandwidth
expansion parameter $k$ to parameters of the system and dataset.  We
validate that theorem empirically for a 30 million-edge prefix of Dataset 3
by analyzing a single aging event that is triggered at \XStream tick $10^7$.  In Figure~\ref{fig:experiment2}, the capacity of one
\XStream processor ($s$) is fixed for each value of of $c$ such that the system will be completely full at the end of the 30 million-edge stream, after processing one aging event\footnote{In particular, if we age at \XStream tick $10^7$ such that fraction $f$ of the edges stored so far survives, then our total storage needed is $S= 20^7\cdot u+10^7\cdot f\cdot u$ and $c = (f\cdot u\cdot 10^7)/S$ 
}.
The number of \XStream processors ($p$) is fixed at $10$. 
We vary over a range of values of $c$ (the target fraction of edges that 
survive the aging event) and $d$ (the
fraction of \XStream ticks in which queries are enabled), we show using
a 3-D surface the predicted $k$ by \ref{thm:infinite-runs}.  We overlay
empirical results in the form of observed data points $(c,d,k)$ 
from \XSCC runs when the bandwidth expansion factor $k$ is set as predicted
by the theorem.  We claim that the prediction surface and observed data 
points corroborate the theorm in this experiment.

\subsection{Experiment 3: \XSCC runs of arbitrary length}

The most important contribution of our work is our set of ideas regarding
infinite graph streams and running  \XSCC for indefinite periods of time
without filling up or failing.  We corroborate these ideas empirically
in this section for a simple use case: the aging predicate is a simple
timestamp comparison: delete all edges older than \XStream tick $t_a$.
This strategy could be adapted to accommodate other aging predicates.

The primary challenge facing an \XStream system administrator is when to
initiate an aging event and what threshold $t_a$ to use. In this section
we present an automated solution.  The system administrator initializes
the system with a target $c$ value (the fraction of edges that should
survive an aging event).  Then we run \XSCC with the following
aging-invocation protocol.  This could be specified in detailed pseudocode,
but in the interest of space we describe it informally below.

When the tail processor begins to fill, we begin an automated binary
search to find a timestamp $t'$ that will hit our fractional target $c$
of edges that survive the aging event.  We augment the
data structures of each \XStream ring processor to include a small
reservoir of 100 edges, and ensure that this is a representative
sample by using the classical technique
of \emph{reservoir sampling}~\cite{vitter1985random}.  

The binary search proceeds by varying $t'$ between the oldest timestamp
and newest timestamps in the system.  At 
each candidate value of $t'$, each ring processor estimates
the number of edges that will survive an aging event with threshold $t'$.
In one circuit of the ring in a payload bundle, the tail processor will
know whether $t'$ needs to be increased or decreased in order to hit
the target $c$ value.  In a logarithmic number of passes, the tail
processor knows an accurate value of $t'$ and tells the head to initiate
aging with that threshold.  In practice, this should give plenty of
time to complete aging honoring Theorem~\ref{lemma:aging-lead-time}.

Figure~\ref{fig:experiment3} depicts the result of running \XSCC on 
a 300 million-edge prefix of Dataset 3 using this automated aging
strategy with a target $c$ value of 0.5.  We see that the binary 
search succeeds in finding aging thresholds that reliably 
reestablish a storage level of $cS$ over an arbitrary number of
aging events (we depict the first 27).

%

\section{Non-constant queries and commands} \label{sec:non-constant}
As we have shown, connectivity queries propagate through the \XStream ring
processings in $p$ \XStream ticks, and the query answer is sent from the
tail processor to the head, then back to the I/O processor. Another potentially
useful query that finishes in $p$ \XStream ticks is ``How many edges are in the system?''
\Cindy{Add some more here.  It can just be a list. Many that one might think should be constant probably require some data structure support, in whcih case, add them at the end.} \Cindy{Revisit notation.}

\XStream also supports queries with non-constant-sized output. At most
one such query can be active at a time. The answer to the query is
output in constant-sized pieces using the payload slots. The canonical
non-constant query is a request to output all vertices in small
connected components. Specially, the answer is the names of all
components with at most $\lambda$ vertices and the list of vertices
within them.  This query makes practical sense only in graphs that
have a giant connected component, but most real graphs have one. We
describe how \XStream executes this specific query.

For a local component with name $\eta$, let $s_{\eta}$ be the number
of vertices in $\eta$.  A processor can compute the size of a local
component as the sum of the number of vertices in each building block.
This is $1$ for a primitive building block.  For this discussion, we
assume processors keep track of the number of primitive building
blocks for each local component while building these components. This
adds only constant work per union-find operation.  However, it's also
possible to inialize local-component sizes to zero and compute them
on-the-fly for this query.  But then, the processor does at most k-1
work counting primitive building blocks or outputing the messages
below, which will further delay the query response. Processors receive
the size of non-primitive building blocks from upstream processors.

When the head processor receives the query ``Output the vertices in
components that have at most $\lambda$ vertices'' in the primary slot
of a bundle, it passes the query downstream in the primary slot. This
allows all processors to learn the type of query and the parameter
$\lambda$. The head then uses the $k-1$ payload slots to start
answering the query.  The query is answered in two phases.  In the
first phase, processors compute component sizes.  For each local
component with name $\eta$, such that $s_{\eta} \le \lambda$, the head
processor (eventually) sends a message ``($\eta, s_{\eta})$'' in a
payload basket. The head outputs $k-1$ of these messages per bundle if
it already knows its component sizes. After the last message, it
outputs a ``query phase done'' token.

Each downstream processor passes the initial query downstream.  Then
for each message $(\eta, s_{\eta})$, the processor checks to see if
$\eta$ is a building block for one of its local components $\eta'$.
If it is, then it increments the size of $\eta'$.  If $\eta$ is not a
local building block, the processor sends the message downstream.
When the processor receives the ``query phase done'' token, it knows
the size of all its non-primitive building blocks, and hence knows
the size of all of its local components.  It sends its own ``($\eta,
s_{\eta})$'' messages for each local component $\eta$ such that
$s_{\eta} \le \lambda$. When it has sent all its messages, it passes
the ``query phase done'' token downstream. If the current graph has a
connected component $\eta_G$ that has size at most $\lambda$, the
message with its final size is passed through the tail and out to the
analyst.  The tail also passes the ``query phase done'' token to the
head.

Sealed processors (full of tree edges) can set a flag indicating they
have computed their component sizes. If there is another such query
before an aging, then it removes messages associated with its local
building blocks without incrementing any size counters.

In the second phase, the head processor (eventually) sends a message
$(\eta, v_i)$, for each primitive vertex $v_i$ in each local component
$\eta$ reported in the first phase.  For the head, all building blocks
are primitive vertices. It's possible to put more than one vertex in
the latter kind of message (e.g. $(\eta, v_1, v_2, v_3)$, depending
upon the size of a slot.  After the last such message, the head passes
a ``query done'' token downstream.

When a downstream processor receives a message $(\eta, v_i)$ from
upstream in the second phase, it checks to see if $\eta$ is a building
block for one of its local components $\eta'$. If not, then it passes
the message downstream.  If so, then $s_{\eta'} \le \lambda$ (i.e. the processor
reported local component $\eta'$ in the first phase, it relabels the message,
sending $(\eta', v_i)$ downstream.  If $\eta'$ is too large, it just removes
the message from the system.

When a downstream processor receives the ``query done'' token, it outputs
messages $(\eta, v_i)$, where $\eta$ is a local componet with $s_{\eta} \le \lambda$
and $v_i$ is a primitive building block (vertex) in local component $\eta$.

A somewhat easier non-constant query is spanning tree. Starting with the head, each
processor outputs its tree edges.

Some queries can be either constant-size (latency $p$) or non-constant
depending upon what additional data structures the processors
maintain.  One example is ``What is the degree of node $v$?'' Suppose
each processor maintains adjacency lists for the subgraph it
holds. Then the processor can find the number of edges adjacent to a
vertex $v$ in constant time, given a hash table to access the
adjacency list for each vertex.  In this case, the vertex-degree query
has latency $p$.  The query makes one pass around the ring with the answer progressing
one processor per tick.  Otherwise, without this data structure, each processor will need time to
compute the number of edges it holds that are adjacent to vertex $v$.
In this case, it is a non-constant query.  The message still touches
each processor once, but the processor may require multiple ticks to
compute the number to add to the accumulating degree value.

Linear algebraic computations typically involve a matrix-vector product,
which would be unweildy to compute directly in the \XStream model.
However, the emerging field of randomized linear
algebra~\cite{drineas2018lectures}
offers a path forward. If we devote some space in the tail processor to
accommodate a sample of edges (adjusting Lemma~\ref{lemma:aging-lead-time}
accordingly), payload slots can be used to accumulate a random sample
of the graph.  Techniques such as randomized 
PageRank~\cite{gasnikov2015efficient} or others might then be applied in a 
separate
thread in the tail processor, still with minimal interruption to the input 
stream.

\section{Conclusion and Future work}
We have provided the first comprehensive set of ideas to handle
infinite graph streaming with bulk expiration events, including theory
and a prototype implementation.  Despite its use of the network interconnect
and MPI message passing rather than memory access, the prototype sometimes
matches the ingestion rate of a purely on-node Intel/TBB benchmark.  Slow
downs are data-dependent and might be mitigated by a multithreaded
implementation and algorithm engineering.  Furthermore, performance of
a single \XSCC ring will benefit from advances in computer architecture.
Although our prototype operates correctly, future work would be
necessary to engineer a production version.

To close, we consider the possibility improving X-Stream's ingestion
rate by orders of magnitude.  It is
possible to do this if we leverage a key property of most real graphs:
\emph{the giant component}.  Suppose that we must ingest such a graph
via hundreds or thousands of disjoint streams, and suppose that we
instantiate an independent \XSCC instance for each.  We note that with
overwhelming likelihood, each \XSCC instance will ingest a portion of
the global giant component.  Using ideas from Section~\ref{sec:non-constant},
each \XSCC instance can stream its $O(\log n)$-sized components out to
a ``small-component server'' (and notify that server of vertices in
components that have joined the giant component).  The small-component
server would handle any connectivity query not involving the giant
component (of which there are relatively few). Full detail is
beyond the scope of this paper and we leave for future work.

\begin{acknowledgements}
Sandia National Laboratories is a multimission laboratory managed and operated by National Technology \& Engineering Solutions of Sandia, LLC, a wholly owned subsidiary of Honeywell International Inc., for the U.S. Department of Energy’s National Nuclear Security Administration under contract DE-NA0003525.
This research was funded through the Laboratory Directed Research and Development (LDRD) program at Sandia.
This paper describes objective technical results and analysis. Any subjective views or opinions that might be expressed in the paper do not necessarily represent the views of the U.S. Department of Energy or the United States Government.
We thank Siva Rajamanickam, Cannada Lewis, and Si Hammond for useful
discussions and baseline code for the TBB benchmark.
\end{acknowledgements}

%
%

\bibliographystyle{plain}      
\bibliography{ms}

\end{document}